\documentclass[pdflatex,sn-mathphys-num]{sn-jnl}
\usepackage{amsmath,comment}
\usepackage{amssymb}
\usepackage{amsfonts} 
\usepackage{soul,booktabs}

\usepackage[footnotesize]{subfigure}

\renewcommand{\eqref}[1]{(\ref{#1})}

\newcommand{\eps}{{\displaystyle \varepsilon}}
\newcommand{\bsub}{\begin{subequations}}
\newcommand{\esub}{\end{subequations}$\!$}

\newcommand{\bigoh}{\mathcal{O}}

\newcommand{\braket}[1]   {\left<{#1}\right>}
\newcommand{\bx}{\mathbf{x}}

\newcommand{\dd}{\textrm{d}}
\newcommand{\R}{\mathbb{R}}

\newcommand{\al}[1]{\textcolor{black}{#1}}

\newcommand{\sdl}[1]{\textcolor{black}{#1}}

\newcommand{\J}{{\mathcal{J}}}

\newcommand{\E}{{\mathbb{E}}}

\renewcommand{\P}{\mathbb{P}}

\newcommand{\erfc}{\mathop{\rm {erfc}}}

\begin{document}

\title[Directional sensing via extreme first passage events.]{Single-cell directional sensing at ultra-low chemoattractant concentrations from extreme first-passage events}

\author[1]{\fnm{Vincent} \sur{Fiorino}}\email{vfiorino@nd.edu}

\author[2]{\fnm{Sean D.} \sur{Lawley}}\email{lawley@utah.edu}

\author*[1]{\fnm{Alan E.} \sur{Lindsay}}\email{a.lindsay@nd.edu}

\affil*[1]{\orgdiv{Department of Applied and Computational Mathematics and Statistics}, \orgname{University of Notre Dame}, \orgaddress{ \city{Notre Dame}, \postcode{46656}, \state{IN}, \country{USA}}}

\affil[2]{\orgdiv{Department of Mathematics}, \orgname{University of Utah}, \orgaddress{ \city{Salt Lake City}, \postcode{84112}, \state{UT}, \country{USA}}}

\abstract{
We investigate single-cell directional sensing from diffusing chemoattractant signals released by a localized source. We focus on the low-concentration regime in which receptor activity is discrete and cellular decisions are made on timescales far shorter than those required for steady-state concentration profiles or receptor saturation to emerge. We derive analytic expressions for the joint distribution of receptor binding times and binding locations, conditional on the position of the source. We show that early binding events carry disproportionately more information about source directionality than later arrivals. Motivated by this observation, we propose and analyze several source-localization estimates that exploit early receptor binding statistics. \al{For a directional estimate formed from an equally weighted average of the earliest impacts, we demonstrate that this estimate is accurate even from a small number of impacts, and derive an expression for the number of binding events that minimizes the mean squared error.} Our results demonstrate that cells possess sufficient information to rapidly and accurately infer the directionality of sparse signals from a localized source through a small number of discrete receptor engagements, well before a steady-state gradient has been established.}


\maketitle

\section{Introduction}

The aim of this paper is to explore the theoretical limits by which cells can establish the directionality of chemical cues at ultra-low chemoattractant concentrations. A key cellular function is the ability to rapidly and accurately interpret the directionality of chemical or electric gradients to inform decisions on where to move, when to divide or whether to initiate a response to a threat \cite{Copos2025,Lew2019,Parent1999,Mogilner2023,NwogbagaCamley2024,Camley2025}. The information regulating these decisions is encoded by the noisy binding activity of diffusing \al{signals} at membrane bound receptors  \cite{Camley24,Mugler2016,Camley2018,berg1977,aquino2016,fancher2017}.

The ability to orient towards a source has been demonstrated across many model systems, including yeast \cite{Lew2019,Lakhani2017,Ismael2016}, Dictyostelium \cite{Parent1999} and neutrophils \cite{Millius2009,Servant2000,Levchenko2002}. In the case of chemotaxis in neutrophils, numerous experimental studies \cite{Millius2009,Orion1999,Servant2000} have demonstrated that cells rapidly identify the directionality of an instantaneous chemoattractant source (administered by pipette). For example, neutrophils demonstrate asymmetric recruitment of membrane components within five seconds of exposure to a $10\mu$M concentration of the chemoattractant FMLP presented by a micropipette at around two cell radii distance \cite{Orion2002,Orion1999}. The timescale of such a response is well before a steady-state gradient is established \cite{LindsayRSOS2023}.

\begin{figure}
\centering
\includegraphics[width = 0.9\textwidth]{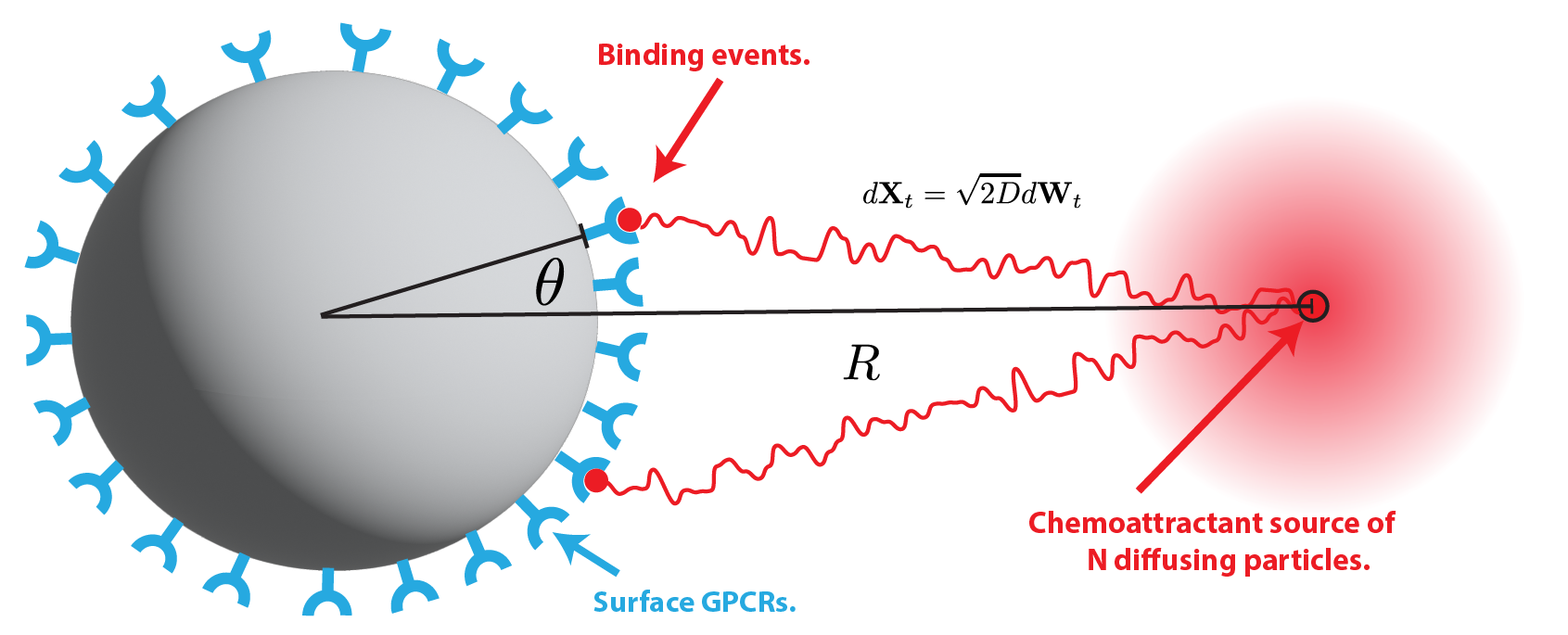}
\caption{Schematic of source detection from the discrete arrival of signaling molecules to a circular cell. At $t=0$, a source located $R>1$ cell lengths away emits $N$ diffusing molecules of diffusivity $D>0$. These signaling molecules diffuse freely before being absorbed at a surface receptor (GPCR). The extreme directional sensing problem is the following: given arrival times $\{T_{j,N}\}_{j=1}^k$ and corresponding polar locations $\{\theta_{j,N}\}_{j=1}^k$ of the first $k$ impacts based on $N$ initial walkers, what estimates can the cell form for the directionality of the source? \label{fig:intro}}
\end{figure}

A commonality across these systems is the role of membrane bound G protein-coupled receptors (GPCRs) which bind to signaling molecules \cite{Boltz2022} and transmit information to downstream signaling cascades. These \lq\lq sensors\rq\rq\ therefore act as intermediaries between extracellular chemoattractants and downstream pathways which either inhibit or initiate necessary cellular responses.

Previous experiments utilized chemoattractant concentrations which saturated GPCRs. However, as the precision of experimental measurement has expanded, it is now clear that  GPCRs can be activated at remarkably low levels of chemoattractant concentrations \cite{civciristov2018}, including at the femtomolar ($10^{-15}$ M) and attomolar ($10^{-18}$ M) levels \cite{Civciristov2019,CALABRESE2021}. At such low concentration levels, receptor engagement and consequently cellular signaling decision making, are regulated by a small number of discrete binding events governed by extreme first passage time statistics. This work expands upon several recent studies that have employed extreme first passage time theory to understand decision making processes in biological systems \cite{Linn2024,LM2023,LawleyJohnson2023,newby2016}.

The aim of this paper is to give a mathematical analysis of the directional information that is encoded in signals associated with such low concentrations of chemoattractant. To this end, we consider an idealized non-dimensional scenario akin to the experiments of \cite{Orion2002,Orion1999} where a circular cell (normalized to unit radius) is centered at the origin while a source of diffusing chemoattractant is placed at $\bx_0 = R[\cos\theta_0,\sin\theta_0]$ for $R>1$ (see Fig.~\ref{fig:intro}). The quantity $R$ represents the multiple of the cell radius at which the source is placed. Intuitively, we expect that detection at larger values of $R$ is more challenging than source detection for values $R\approx 1$.  A key element of our approach in the low chemoattractant limit is to consider the signaling molecules to be discrete. \al{In this setting we assume at the initial time $t=0$ that $N$ molecules are released from the source and undergo unbiased random motion before eventually reaching a receptor on the cell surface.} As an example of typical values, a release of $1$mL of chemoattractant at femtomolar to attomolar concentrations would represent a release of approximately $N\approx 10^3-10^6$ molecules. In our analysis we will assume a dense covering of GPCRs so that a continuum approximation is valid.

\al{Each of the $N$ molecules released at the source eventually reaches the cell surface and binds a receptor, producing two observable quantities: the time of the binding event and its location on the cell surface. Directional sensing is then an inverse problem: from the earliest binding events, what can the cell infer about the source? We will show that the two observables carry complementary information. The arrival times reflect the source distance, while the arrival locations reflect the source direction. We introduce the formal notation for the sets of arrival times and locations in Section 2, where the extreme-value statistics of the earliest arrivals are developed.}

The most pertinent biological information on the source is its direction relative to the cell. It was noticed in \cite{LindsayRSOS2023} that when recovering the location of a diffusive source from the time-dependent signal, the short time signal ($0<t\ll1)$ gave a strong estimate on the directionality of the source but yielded poor information on the distance. Heuristically, this is because molecules which are quickly absorbed at the cell surface have likely taken very direct or near-ballistic paths. This idea was further explored in \cite{LindBJ2023} which showed in a three dimensional setting that the directional information is contained primarily in a small fraction of the earliest signals received by the cell. Impacts that arrive after sufficiently long sojourns have largely lost information about their initial location and therefore dilute the directional information carried by earlier arrivals. This observation highlights a challenge in recovering source direction from equilibrium quantities (such as splitting probabilities), since such estimators necessarily average over signals of lower informational quality \cite{LLM2020,Dobramysl2018a,Dobramysl2018b}. By using only the early fraction of the signal, it was shown in \cite{LindBJ2023} that a simple average of just a few impact locations can provide an accurate directional estimate.  Here we build on these studies by deriving the joint statistics of early receptor binding times and impact locations in 2D. 

\al{We find that the information a cell can extract about the source separates into two channels: the times of early binding events, which encode the source distance $R$, and the locations of those events, which encode the source direction $\theta_0$. Section 3 asks how accurately the source distance can be inferred from arrival times alone, using maximum-likelihood estimation and the associated Fisher information both to quantify the best achievable accuracy and to compare candidate estimators. Sections 4–5 turn to the directional problem, deriving the distribution of arrival locations and analyzing estimators of source direction. In both cases we quantify the achievable accuracy as a function of the number $k$ of early binding events.}

\al{The main mathematical result of this work is a  derivation of the limiting distribution of the first $k$ arrivals in this extreme statistics regime.} Specifically, in Sec.~\ref{sec:2d_arrivals}, we find that the impact location of the $k^{\mathrm{th}}$ arrival to a circular cell has the limiting behavior
\begin{equation}\label{intro_arrival_k}
\theta_{k,N} \sim \mathcal{N}(\theta_0,\sigma^2_{k,N}), \qquad \sigma^2_{k,N} = \frac{2D}{R}\Big(b_N + a_N \log k \Big),
\end{equation}
as $N\to\infty$. Here $a_N$ and $b_N$ are quantities found in terms of the limiting properties of the survival probability for a  single particle which we also obtain. For the specific case of an all absorbing circular cell of unit radius, the leading order limiting behavior is found to be
\begin{equation}\label{intro_arrival_k_limit}
\theta_{k,N} \sim \mathcal{N}(\theta_0,\sigma^2_{k,N}), \qquad \sigma^2_{k,N}  \sim \frac{(R-1)^2}{RW}\left( 1 + \frac{2\log k}{1+W}\right), \qquad W = W_0\Big( \frac{2N^2}{\pi R} \Big),
\end{equation}
as $N\to\infty$. This behavior is based on the limiting forms $a_N\sim (R-1)^2/(DW(1+W))$ and $b_N\sim(R-1)^2/(2DW)$ as $N\to\infty$. Here $W_0(z)$ is the principal branch of the LambertW function \cite{corless1996}, which has behavior $W_0(z) = \log z - \log \log z+ \cdots$ as $z\to\infty$. Hence $W_0\sim 2\log N$ as $N\to\infty$. The result reveals the limits on the accuracy of directional inference based on the first few early arrivals together with the dependence on key parameters, including source distance $(R)$, number of early arrivals $(k)$ and the strength of the initial signal $(N)$. To our knowledge, this is the first derivation and analysis of the joint asymptotic distribution of early hitting times and impact locations for diffusive capture in two dimensions.

\al{A surprising consequence of \eqref{intro_arrival_k_limit} is that the breadth of angular space sampled by the early arrivals is independent of the single particle diffusivity $D$. We will demonstrate that a larger diffusivity delivers the $k^{\mathrm{th}}$ impact proportionally sooner by exactly the factor by which it broadens the angular distribution. Hence the parameter $D$ sets only the overall timescale of capture, not the geometry of the early-impact distribution.}

Using the detailed statistical properties of the early arrivals, we investigate the accuracy of several estimates that a cell could form to infer the source direction. We find that the impact times largely inform on the source distance while the impact locations encode the directionality. Our analysis shows that a cell can employ simple averaging over the first $k$ arrivals to reduce the variance (error) in the accuracy.

\al{Maximum likelihood estimation (MLE)} can be used to form more accurate estimates of the source \cite{EndresWingreen2009}. However, implementation of these estimators necessitates additional computational capabilities, such as the solution of nonlinear equations. Our analysis is able to quantify the accuracy in a range of estimators, from simple averages to relatively sophisticated MLEs. As an example of our theory, we analyze the simple average of the first $k$ impacts given by $\textbf{n}^{\text{res}} = \frac{1}{k}\sum_{j=1}^k (\cos\theta_j,\sin\theta_j)$. We find that the error $\rho^{\text{res}}_{k} = 1-\textbf{n}^{\text{res}}\cdot (\cos\theta_0,\sin\theta_0)$ associated with this estimate has mean and variance
\bsub\label{eqn:error_avg}
\begin{align}
\label{eqn:error_avg_a} \mathbb{E}[\rho_k^{\text{res}}] &\approx \frac{D}{R}\Big(b_N + a_N \big(\log k -1\big)\Big), \\
\label{eqn:error_avg_b} \text{Var}[\rho_k^{\text{res}}] & \approx \frac{2D^2}{R^2k}\Big( \big(a_N \log k + b_N -a_N \big)^2 + a_N^2 \Big).
\end{align}
\esub
\al{The error \eqref{eqn:error_avg_a} grows with $k$ since the first impact has the lowest error while subsequent individual arrivals have logarithmically growing error. However, while the error increases in $k$, the variance in the error \eqref{eqn:error_avg_b} decreases. These opposing trends are a bias–variance trade-off. By considering the mean squared error $\mathbb{E}[(\rho_k^{\text{res}})^2]$, we show that this trade off results in an optimal $k = k^{\ast}$ which satisfies a transcendental equation. As $N\to\infty$, this critical number has asymptotic behavior $k^{\ast} \sim \log N$. For $k>k^{\ast}$, additional impacts produce a larger error in the resultant $\textbf{n}^{\text{res}}$. Hence an accurate estimate of the source direction can be formed from relatively few impacts $k$.}

The remainder of the article is structured as follows. In Sec.~\ref{sec:derivation_limiting} we describe preliminary results on the distributions of extreme arrival times, including those of individual arrivals and the joint distribution of $k$ arrivals $\mathcal{T}_N$ and their differences. We emphasize that it is necessary to first describe the statistics of arrival times before arrival locations since the former depends on the latter. In Sec.~\ref{sec:MLE_time} we introduce and analyze several estimates of the source distance using only the arrival times. These quantities primarily encode information on the distance of the source. In Sec.~\ref{sec:2d_arrivals}, we derive the distribution \eqref{intro_arrival_k} of angular arrivals to a circular cell in the case of both Dirichlet and Robin boundary conditions. In Sec.~\ref{sec:Source_Extreme}, we use the result \eqref{intro_arrival_k} to investigate the accuracy of several estimators of the source direction.
Finally, in Sec.~\ref{sec:discussion} we discuss future avenues of investigation that arise from this study.
 

\section{Statistics of extreme first passage times.}\label{sec:derivation_limiting}

In this section, we introduce key results on the distribution of the set of extreme hitting times which have been a topic of significant recent research in the literature \cite{Linn2024,Linn_2022,MacLaurin2025,Tung2025}. It is necessary to characterize the hitting times first before the hitting locations. This is due to the fact that particles arriving quickly to the cell have hitting distributions tightly focused on the source direction, while particles that undergo longer sojourns are distributed more evenly over the cell surface.

\al{We denote the arrival times of the $N$ molecules by $\{T_{j,N}\}_{j=1}^N$ and their arrival locations on the circular cell by $\{(\cos \theta_{j,N}, \sin \theta_{j,N})\}_{j=1}^N$, where the index $j = 1,\ldots,N$ labels the molecules and $N$ is the number released.}

For an independent and identically distributed (i.i.d.)~sequence of first passage times $\{\tau_1,\ldots,\tau_N \}$ with survival probability $S(t):=\mathbb{P}(\tau_1>t)$, \al{the first arrival is defined as
\[
T_{1,N}:= \min\{\tau_1,\ldots, \tau_N \}.
\]
The i.i.d. property of the times means that
\[
\mathbb{P}(T_{1,N}>\tau) = \mathbb{P}(\tau_1>\tau) \times \cdots \times \mathbb{P}(\tau_N>\tau)  = [S(\tau)]^N.
\]
For $N\gg1$ and $0\leq S(\tau)\leq1$, the mass becomes concentrated near $t=0$. Hence, describing extreme statistics requires detailed knowledge of the short time asymptotic behavior of $S(t)$ as $t\to0^{+}$. For a large class of diffusion problems in various dimensions, geometric settings and boundary conditions \cite{lawley2020networks,Linn_2022} this behavior can be mapped onto the family of functions }
\begin{equation}\label{eqn:main_Surv}
    1-S(t) \sim At^p e^{-\frac{C}{t}}, \quad \mbox{as} \quad t\to0^{+}.
\end{equation}
Here $A$, $C$, $p$ are constants with $A$ and $C$ being strictly positive. \al{A simple example we will make extensive use of is diffusion on the half line with release at $x=R$ and capture at $x=0$. The survival probability is $S(t) = 1 - \erfc(\frac{R}{\sqrt{4Dt}})$ and the limiting behavior as $t\to0^{+}$ is given by 
\[
1 - S(t) \sim \sqrt{\frac{4Dt}{\pi R^2}} e^{\frac{-R^2}{4Dt}}, \qquad \mbox{as} \qquad t\to0^{+}.
\]
}
More generally, the $k^{\mathrm{th}}$ fastest arrival time at the cell is defined as
\bsub\label{eqn_limitingT}
\begin{equation}\label{eqn_limitingT_a}
    T_{k,N} := \min\big\{ \{\tau_1,\ldots, \tau_N\} \setminus \cup_{j=1}^{k-1} \{T_{j,N}\}  \big\}, \qquad k \in\{1,\ldots,N \}.
\end{equation}
Refined statistics on the distributions of the $k^{\mathrm{th}}$ arrival $T_{k,N}$ have recently been determined in the limit as $N\to\infty$. A key result \cite[Thm.~4]{lawley2020dist} utilized in the present work is the following convergence in distribution,
\begin{equation}\label{eqn_limitingT_b}
    \left(\frac{T_{1,N}-b_N}{a_N},\ldots,\frac{T_{k,N}-b_N}{a_N} \right)\to_\dd  \textbf{X}^{(k)} := (X_1,\ldots, X_k)\in\mathbb{R}^k, \quad \mbox{as} \quad N\to\infty.
\end{equation}
\esub
Here the joint probability density function of $\textbf{X}^{(k)}\in\mathbb{R}^k$ is
\begin{equation}\label{eqn:dist_full}
    f_{\textbf{X}^{(k)}}(x_1,\ldots,x_k) = \left\{ \begin{array}{rc} \exp(-{e^{x_k}}) \prod_{r=1}^k e^{x_r}, \quad &x_1\leq \cdots \leq x_k;\\[4pt]
    0, \quad & \mbox{otherwise.}\end{array}\right.
\end{equation}
\al{The scale parameters $a_N$ and $b_N$ set the distribution of the early arrival times in the limiting law \eqref{eqn_limitingT}.} The constants $a_N$ and $b_N$ are defined as
\bsub\label{eqn:AandB}
\begin{align}
\label{eqn:AandB_a}   b_N = \frac{C}{\log(AN)}, \quad  a_N &=\frac{b_N}{ \alpha }, \qquad  \alpha  = \log (AN) , &\mbox{if} \quad p = 0;\\[4pt]
\label{eqn:AandB_b} b_N = \frac{C}{pW},\quad a_N &= \frac{b_N}{\alpha},  \qquad \alpha = p(1+W), & \mbox{if} \quad p \neq 0.
\end{align}
\esub
\al{We will show these quantities are defined in terms of the source distance and diffusivity so that estimating $a_N$ from the arrival times is therefore equivalent to estimating the source distance $R$ for fixed $D$.} The constant $W$ is defined as
\begin{equation}\label{eqn:FormW}
    W = \left\{ \begin{array}{rl}
    W_0\big( (C/p) (AN)^{1/p} \big), & \quad p>0,\\[5pt]
    W_{-1}\big( (C/p) (AN)^{1/p} \big), &\quad p<0,
    \end{array}\right.
\end{equation}
where $W_0(z)$ and $W_{-1}(z)$ denote the principal and lower branches of the LambertW function respectively. \al{The principal branch of the LambertW function has large argument asymptotic representation \cite{corless1996}
\begin{equation}
    W_0(z) = \log z - \log \log z + \mathcal{O}\left(\frac{\log \log z }{\log z}\right) + \cdots \quad \mbox{as} \quad z \to\infty.
\end{equation}
In the case $p>0$, we will often invoke the leading order behavior $W\sim \frac1p\log N$ as $N\to\infty$ to obtain expressions that are less accurate but offer more biological interpretability. To retain qualitative accuracy at moderate values of $N$, the full form described in \eqref{eqn:FormW} is retained.}

\begin{figure}
    \centering
    \subfigure[Convergence of $T_{1,N}$ to the density \eqref{eqn:T_dist}.]{\includegraphics[width=0.475\textwidth]{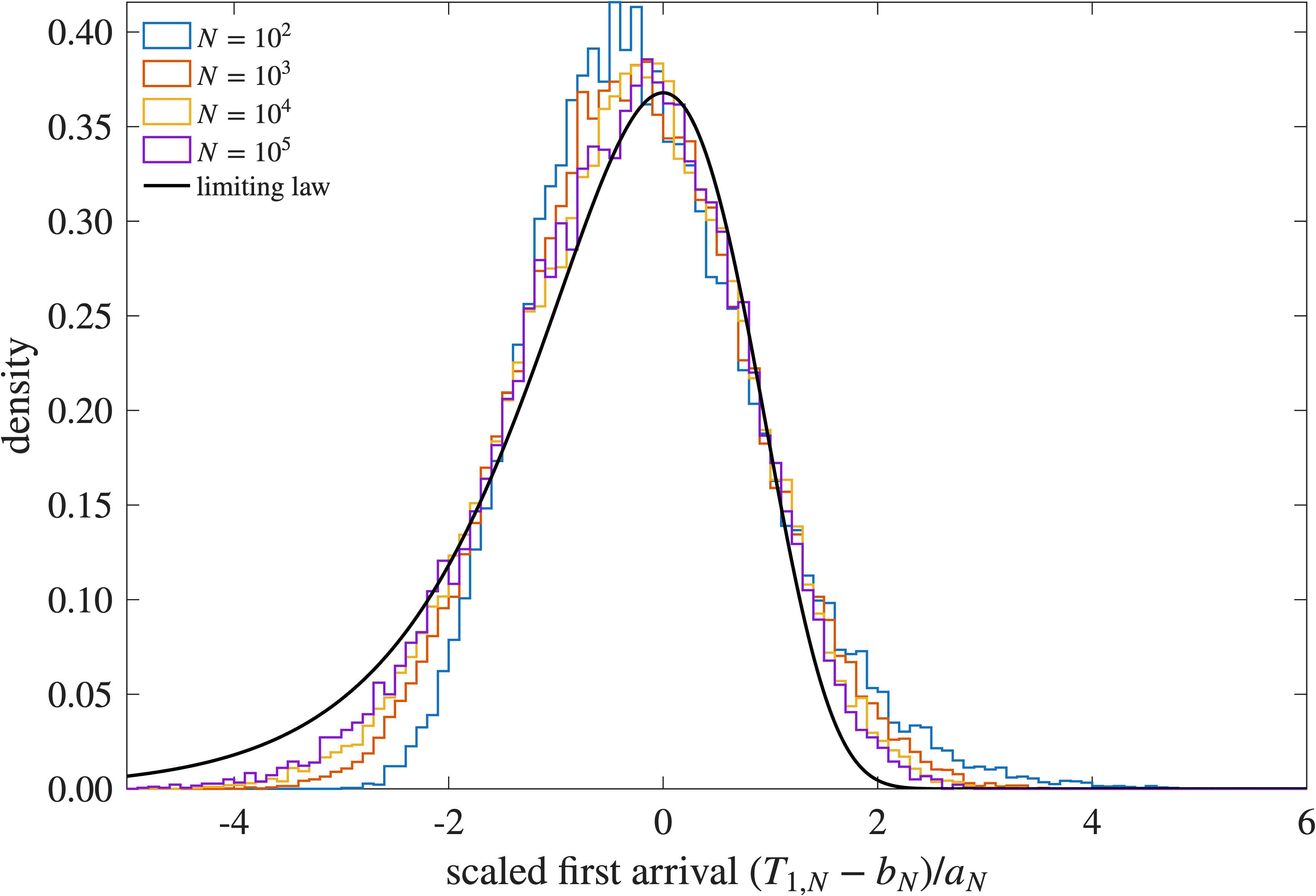}
    \label{fig:convergence_gumbel_a}}\quad
    \subfigure[Spread of first impact $\sigma_{1,N} = \sqrt{2D b_N/R}$.]{\includegraphics[width=0.475\textwidth]{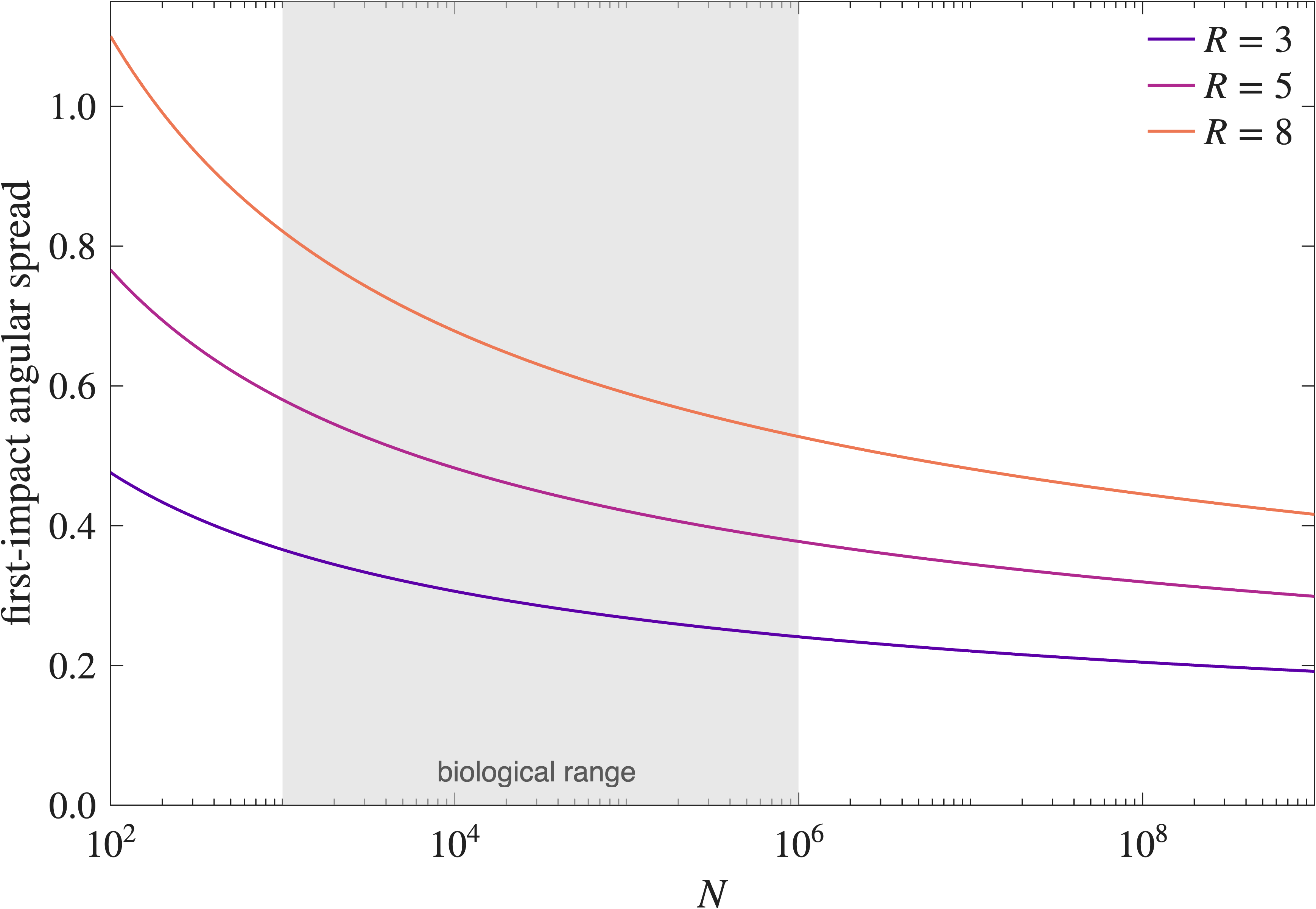}
    \label{fig:convergence_gumbel_b}}
    \caption{\al{Convergence to the limiting distribution as $N\to\infty$. In Panel (a), we show convergence of empirical distributions of $(T_{1,N}-b_N)/a_N$ to the limiting distribution $f_{X_1}(x) = \exp(x-\exp(x))$ for a range of $N$. In panel (b), we plot the angular spread \eqref{intro_arrival_k} of the first arrival $\sigma_{1,N} = \sqrt{2Db_N/R}$ to the cell. Over several decades in $N$ we observe that the limiting distribution accurately captures the low concentration limit and that the angular spread varies modestly, hence providing a robust measure of directional sensing capabilities.} \label{fig:convergence_gumbel}}
\end{figure}

For the individual arrival times, we have that the $k^{\mathrm{th}}$ fastest arrival has the limiting behavior 
\begin{equation}
    \frac{T_{k,N}-b_N}{a_N} \to_\dd X_k, \quad \mbox{as} \quad N\to\infty,
\end{equation}
where $X_k$ has the probability density function
\begin{equation}\label{eqn:T_dist}
    f_{X_k}(x) = \frac{\exp(kx-e^x)}{(k-1)!}.
\end{equation}
In particular, we have that (Theorem 5 of \cite{lawley2020dist})
\bsub\label{eq:ExpTime}
\begin{align}
\label{eq:ExpTime_a}  \mathbb{E}[T_{k,N}]&= b_N + \psi(k)a_N + o(a_N),\\[4pt]
\label{eq:ExpTime_b}    \mbox{Var}(T_{k,N}) &= \psi'(k) a_N^2 + o(a_N^2),
\end{align}
\esub
where $\psi(k)= H_{k-1} - \gamma_e$ is the digamma function, $H_k = \sum_{r=1}^k\frac{1}{r}$ is the $k^{\mathrm{th}}$ harmonic number, and $\gamma_e \approx 0.577$ is the Euler-Mascheroni constant. \al{In Fig.~\ref{fig:convergence_gumbel_a} we show Monte-Carlo simulations that demonstrate the convergence of $T_{1,N}$ to the limiting distribution \eqref{eqn:T_dist} for $R=5$ and $D=1$ in the case of the one-dimensional survival probability $S(t) = 1 - \erfc(\frac{R}{\sqrt{4Dt}})$. Over the biologically relevant range of $N$ ($10^3-10^6$), we observe that the two distributions have broad agreement in shape. This is in agreement with related convergence studies performed in \cite{lawley2020dist}. In addition, we plot in Fig.~\ref{fig:convergence_gumbel_b} the quantity $\sigma_{1,N} = \sqrt{2Db_N/R}$ which characterizes the angular extent of the first impact. This quantity was stated in \eqref{intro_arrival_k} and will be derived in Sec.~\ref{sec:2d_arrivals}. The key observation is that this quantity exhibits modest variations in value while the number of source particles $N$ varies over several decades. Hence the biologically relevant range of values for $N$ is well captured in the asymptotic limit $N\to\infty$.}

\al{The full record of arrival times $\{ T_{j,N} \}_{j=1}^k$ is} a set of dependent random variables satisfying the joint distribution \eqref{eqn_limitingT} in the limit as $N\to\infty$. This implies that a cell would need to store a sequence of times in order to process them to reconstruct information on the source. As a consequence, it is natural to ask whether simpler measurements of time can be used for inference. To this end, we derive (see Appendix \ref{app:Times_Diff}) the distribution of the differences between arrival times. For $1\le k\ll N$, we have that the time differences are exponentially distributed, specifically,
\begin{align}\label{eq:exp}
    T_{k+1,N}-T_{k,N}
    \sim_{\dd} \textup{Exp}\left[\frac{k}{a_N}\right],
\end{align}
with mean $a_N/k$ where $a_N$ is given in \eqref{eqn:AandB}. Hence, if we only know the interarrival times, then this value is related to the single parameter $a_N$ which shapes the limiting distribution for $N\gg1$.

\section{Maximum likelihood estimates of source location from extreme time data}\label{sec:MLE_time}

In this section, we describe a maximum likelihood estimation (MLE) process for determining the source parameters from the arrival statistics. \al{The arrival time statistics depend on the source only through $a_N$ which itself depends on $R$ and $D$. Therefore inferring source distance reduces to estimating the single parameter $a_N$. The MLE and its Fisher information that we derive in this section are able to characterize the best achievable accuracy and allow for a systematic comparison of the information content by different summaries of the data: the full arrival times versus the differences.}

\al{To set the stage for our MLE approach, consider a single cellular observation 
of $k$ impact times ${\mathcal{T}}^{(k)} = (T_{1},T_{2},\ldots,T_{k})$ based on $N$ initial walkers. To simplify notation, we omit in this section the dependence of both ${\mathcal{T}}^{(k)}$ and $a= a_N$ on $N$. Suppose we have
$n$ independent and identically distributed realizations
$\{\mathcal{T}^{(k)}_1,\ldots,\mathcal{T}^{(k)}_n\}$ of this observable,
corresponding to repeated releases from a source at the same distance.}

Maximum likelihood estimation infers the unknown parameter $a$ by maximizing
the likelihood of the observed data. For the case of the full set of arrival times, the full likelihood is the product over the
$n$ independent observations,
\begin{equation}\label{eqn:likliehood}
\mathcal{L}(a | \{{\mathcal{T}_j}^{(k)}\}_{j=1}^n) = \prod_{j=1}^{n}f_{ \mathcal{T}^{(k)}}( {\mathcal{T}_j}^{(k)} | a)
\end{equation}
where $f_{ \mathcal{T}^{(k)}}$ is the joint density of the first $k$ arrivals for each observation and given explicitly in \eqref{eqn:dist_full} after rescaling.

The maximum likelihood estimate $\hat{a}$ is
the maximizer of \eqref{eqn:likliehood}. This method is consistent: given a
sufficient number $n$ of observations, it converges in the sense that \cite{vanderVaart1998}:
\begin{equation}\label{MLE_main}
\sqrt{n}\big(\hat{a}- a) \to_\dd \mathcal{N}(0, I^{-1}), \qquad I(a)  = -\mathbb{E} \Big[ \frac{\partial^2}{\partial a^2} \log \mathcal{L}\big(a | {\mathcal{T}}^{(k)}\big) \Big| \ a \Big]
\end{equation}
as $n\to\infty$. The quantity $I(a)$ is the (per-observation) \emph{Fisher information}, a simple quantity that determines the quality of the estimation with larger values corresponding to a more accurate estimate on the correct value. The total information from $n$ independent observations is $nI(a)$.

\al{In the following two sections we build estimators from two different summaries
of this data. In Section~\ref{sec:differences} we use only the interarrival
differences $T_{i+1}-T_i$, which are mutually independent
(Appendix~\ref{app:Times_Diff}) and yield a particularly simple estimator;
in Section~\ref{sec:FullTimes} we use the full arrival times through the joint
likelihood \eqref{eqn:likliehood}. Comparing their Fisher information
quantifies how much information about the source distance is discarded by
retaining only the interarrival differences.}

\subsection{Assuming the cell accesses the time differences}\label{sec:differences}

In the scenario we consider the situation where the time interval between arrivals is exponentially distributed as given in \eqref{eq:exp}.  \al{Let $\{{\mathcal T}^{(k)}_{1},{\mathcal T}^{(k)}_{2},\ldots,{\mathcal T}_n^{(k)}\}$ be a set of $n$ i.i.d. observations for first $k$ arrival times after the release of $N$ molecules at $t=0$. Here ${\mathcal T}^{(k)}_{j} = (T_{1,j},\ldots,T_{k,j})$ where we have assumed the $k$ arrival times are sorted $(T_{i,j}<T_{i+1,j})$. The dependence of $N$ is not explicitly recorded. Under the assumption that the $n$ observations are independent and that, within each observation, the successive interarrival times are mutually independent with $T_{i+1,j} - T_{i,j} \sim_d \text{Exp}(i/a)$ (rate $\lambda_i = i/a$; see Appendix \ref{app:Times_Diff}), the joint probability density (likelihood) of the complete set of interarrival times}
\[
\mathcal{L}\big(a|\{ \mathcal{T}_j^{(k)} \}_{j=1}^n\big) = \prod_{j=1}^{n} \prod_{i=1}^{k-1} \lambda_i e^{-\lambda_i (T_{i+1,j}-T_{i,j})}, \qquad \lambda_i = \frac{i}{a}.
\]
For convenience, the subscript on the unknown parameter $a = a_N$ is suppressed. The negative log of the likelihood is more amenable to analysis and is given by
\[
\ell\big(a|\{ \mathcal{T}_j^{(k)} \}_{j=1}^n\big) = - \log\Big[\mathcal{L}\big(a|\{ \mathcal{T}_j^{(k)} \}_{j=1}^n \big)\Big]
\]
With this definition, we calculate that
\begin{align*}
\ell\big(a|\{ \mathcal{T}_j^{(k)} \}_{j=1}^n\big) &= -\log \prod_{j=1}^{n} \prod_{i=1}^{k-1} \lambda_i e^{-\lambda_i (T_{i+1,j}-T_{i,j})}\\[4pt]
&= \sum_{j=1}^{n}\sum_{i=1}^{k-1} \Big[ \log a + \frac{i}{a} (T_{i+1,j}-T_{i,j}) - \log i\Big]\\[4pt]
&= n(k-1)\log a + \frac{1}{a}\sum_{j=1}^{n} \sum_{i=1}^{k-1} i (T_{i+1,j}-T_{i,j}) - n \sum_{i=1}^{k-1} \log i.
\end{align*}
The first and second derivatives with respect to $a$ yield the score function and the Fisher information respectively. These quantities are given by
\bsub\label{eq:LossAndFisher_Diffs}
\begin{align}
\frac{\partial \ell}{\partial a}
 & = \frac{n(k-1)}{a} - \frac{1}{a^2}\sum_{j=1}^{n} \sum_{i=1}^{k-1} i (T_{i+1,j}-T_{i,j})\\
\frac{\partial^2 \ell}{\partial a^2}& = -\frac{n(k-1)}{a^2} + \frac{2}{a^3}\sum_{j=1}^{n} \sum_{i=1}^{k-1} i (T_{i+1,j}-T_{i,j}).
\end{align}
\esub
The zeros of the score function yield the MLE estimate $\hat{a}$ given by
\begin{equation}\label{eq:dist_est}
\hat{a}_n = \frac{1}{n(k-1)}\sum_{j=1}^{n}\sum_{i=1}^{k-1}i (T_{i+1,j}-T_{i,j}).
\end{equation}
\al{The Fisher information of the full observation set is given by
\begin{equation}\label{eq:diff_Fisher}
 \mathbb{E}\left[\frac{\partial^2 \ell}{\partial a^2} \right] = \frac{n(k-1)}{a^2},
\end{equation}
so that the per-observation information from a single independent observation is
$$I_{\text{diff}}(a) = \frac{k-1}{a^{2}}.$$} To obtain \eqref{eq:diff_Fisher}, we have calculated that
\begin{align*}
\mathbb{E}\left[\sum_{j=1}^{n}\sum_{i=1}^{k-1} i(T_{i+1,j}-T_{i,j}) \right] = na(k-1),
\end{align*}
where from \eqref{eq:ExpTime_a}, we have used that $\mathbb{E}[T_{k,j}]= \mathbb{E}[T_{1,j}]+ H_{k-1}a$. Hence, applying the result \eqref{MLE_main}, we conclude that the estimate \eqref{eq:dist_est} satisfies
\begin{equation}\label{eqn:MLE_error_Diff}
\sqrt{n}(\hat{a}_n - a) \to_\dd \mathcal{N}\Big(0, \frac{a^2}{k-1}\Big), \quad \mbox{as} \quad n \to \infty.
\end{equation}

As a confirmation of the result \eqref{eqn:MLE_error_Diff}, we simulated $n=10^3$ arrival events to create an estimate for $\hat{a}$. This process was repeated over $M=10^4$ trials to build a distribution for these estimates. In Fig.~\ref{fig:FullTimePDF} we show excellent agreement between the theoretically predicted distribution (solid green) and that obtained by simulation (green histogram).

\subsection{Estimation using the full times}\label{sec:FullTimes}
In this section, we describe a scenario where the cell can access the entire set of arrival times. The likelihood function of $n$ independent arrival times $\{T_{1,j},T_{2,j},\ldots,T_{k,j}\}$ for $j=1,\ldots, n$, is given by
\[
\mathcal{L} (a | \{ \mathcal{T}_j^{(k)} \}_{j=1}^n) = \prod_{j=1}^{n}\frac{e^{-e^{x_{k,j}}}}{a^k}\prod_{i=1}^k e^{x_{i,j}}, \qquad x_{i,j} = \frac{T_{i,j} - \alpha a}{a}, 
\]
where $\alpha = p(1+W)$ is assumed constant. The negative log likelihood is given by 
\begin{equation}
\ell = -\log\Bigg(\prod_{j=1}^{n}\frac{e^{-e^{x_{k,j}}}}{a^k} \prod_{i=1}^k e^{x_{i,j}} \Bigg)=nk \log a + \sum_{j=1}^{n}e^{\frac{T_{k,j}-\alpha a}{a}} - \sum_{j=1}^{n}\sum_{i=1}^k \frac{T_{i,j}-\alpha a}{a}.
\end{equation}
The first and second derivatives of $\ell$ are calculated to be
\bsub
\begin{align*}
    \frac{\partial \ell}{\partial a} &= \frac{nk}{a} -\frac{1}{a^2}\sum_{j=1}^{n}T_{k,j}e^{\frac{T_{k,j}-\alpha  a}{a}} + \frac{1}{a^2}\sum_{j=1}^{n} \sum_{i=1}^k T_{i,j},\\[4pt]
    \frac{\partial^2 \ell}{\partial a^2} &= -\frac{nk}{a^2} +\frac{1}{a^4}\sum_{j=1}^{n}(T_{k,j})^2e^{\frac{T_{k,j}-\alpha  a}{a}} +\frac{2}{a^3}\sum_{j=1}^{n}T_{k,j}e^{\frac{T_{k,j}-\alpha  a}{a}} -\frac{2}{a^3} \sum_{j=1}^{n}\sum_{i=1}^k T_{i,j}.
\end{align*}
\esub
The zeros of the score function yield the MLE estimate $\hat{a}$ which is given by
\begin{equation}\label{eq:full_times_nonlin}
0 = \hat{a} -\frac{1}{nk}\sum_{j=1}^n T_{k,j}e^{\frac{T_{k,j}-\alpha   \hat{a}}{ \hat{a}}} +\frac{1}{nk}\sum_{j=1}^n\sum_{i=1}^kT_{i,j}.
\end{equation}
Introducing the change of variable $z$ = $\frac{1}{ \hat{a}}$ reduces \eqref{eq:full_times_nonlin} to
\begin{equation}\label{eq:rescaled_fullTimes}
e^\alpha - \frac{z}{nk}\sum_{j=1}^n T_{k,j}e^{zT_{k,j}} +\frac{z}{nk}e^\alpha \sum_{j=1}^n\sum_{i=1}^kT_{i,j} = 0.
\end{equation}
In this case, the equation \eqref{eq:rescaled_fullTimes} is nonlinear and we utilize a bisection method to access its roots $\hat{a}_n$.
%
The Fisher information is
\begin{align}
\nonumber I(a) &= \mathbb{E}\left[\frac{\partial^2}{\partial a^2} \ell({\mathcal{ T}^{(k)}};a) \Big| a \right] \\[5pt]
\nonumber &= \mathbb{E}\left[\sum_{j=1}^{n} \left(\Big(\frac{T_{k,j}^2}{a^4} + \frac{2T_{k,j}}{a^3}\Big) e^{\frac{T_{k,j}-\alpha a}{a}}\right)\right] -\frac{2}{a^3} \mathbb{E}\Big[\sum_{j=1}^{n}\sum_{i=1}^k T_{i,j}\Big] - \frac{nk}{a^2}\\[5pt]
&= n \underbrace{\mathbb{E}\left[ \left(\Big(\frac{T_{k}^2}{a^4} + \frac{2T_{k}}{a^3}\Big) e^{\frac{T_{k}-\alpha a}{a}}\right)\right]}_{I_1} - \frac{2n}{a^3} \underbrace{\mathbb{E}\Big[\sum_{i=1}^k T_{i}\Big]}_{I_2} - \frac{nk}{a^2}.\label{eqn:FisherFull}
\end{align}
We now proceed to calculate the expectations $I_1$ and $I_2$ separately. 

\paragraph{Calculation of $I_1$:} The probability density function of $T_{k}$, given by \eqref{eqn:T_dist}, is
\[
\mathbb{P}(T_{k} = t) = \frac{e^{k(\frac{t - \alpha a}{a}) - e^\frac{t - \alpha a}{a}}}{a(k - 1)!}.
\]
Hence, by defining the auxiliary functions $h(t)$ and $g(t)$ as
\[
h(t) = \frac{1}{a(k-1)!} \left(\frac{t^2}{a^4} + \frac{2t}{a^3}\right), \qquad g(t) = (k+1)(t-\alpha a) - a e^{(a^{-1}(t-\alpha a))},
\]
we can express the integral $I_1$ as
\[
I_1 = \mathbb{E}\left[ \left(\Big(\frac{T_{k}^2}{a^4} + \frac{2T_{k}}{a^3}\Big) e^{\frac{T_{k}-\alpha a}{a}}\right)\right] = \int_{t=0}^{\infty} h(t) e^{a^{-1}g(t)}\dd t.
\]
Applying Laplace's method for $a\ll1$, we approximate this integral as 
\begin{align*}
I_1 &\approx \sqrt{\frac{2\pi a}{|g''(t_c)|} }h(t_c) \exp[a^{-1}g(t_c)] \\[5pt]
      &= a \sqrt{\frac{2\pi}{k+1}}h(t_c)\exp\Big[\big(k+1\big)\big(\log(k+1)-1\big)\Big]
 \qquad \mbox{for} \qquad a\ll1.
\end{align*}
Here we have calculated
\[
t_c = a [\alpha  + \log(k+1)], \qquad g({t_c}) = a(k+1) [\log(k+1)-1], \qquad 
|g''(t_c)| = \frac{k + 1}{a}.
\]
After some rearrangement and simplification, we obtain that
\begin{equation}
I_1 \approx \sqrt{\frac{2\pi}{k+1}}\left( \frac{t_c^2}{a^4} + \frac{2 t_c}{a^3} \right)  \frac{1}{(k-1)!} \left(\frac{k+1}{e}\right)^{k+1}.
\end{equation}
This expression can be further simplified using Stirling's approximation
\[
(k+1)!\approx \sqrt{2\pi(k+1)}\left(\frac{k+1}{e}\right)^{k+1},
\]
which yields
\begin{equation}\label{eqn:I1_final}
I_1 = \Big( \big(\log(k+1)+ 1 +\alpha \big )^2 -1\Big) \frac{k}{a^2}.
\end{equation}


\paragraph{Calculation of $I_2$:} Recalling from \eqref{eqn:AandB} that $b = \alpha a$ and from \eqref{eq:ExpTime_a} that $\mathbb{E}[T_{k}] \approx a\big(\alpha + \psi(k)\big)$, we calculate
\begin{align*}
I_2 = \mathbb{E}\left[\sum_{i=1}^{k} T_{i} \right] &= k a \alpha  + a\Big[ \psi(1) +  \psi(2) + \cdots + \psi(k)\Big]\\[4pt]
& = ka (\alpha - \gamma_e) + a\left[\sum_{i=1}^{k-1} H_{i} \right] = ka(\alpha - \gamma_e) + a(kH_{k-1} -k+1),
\end{align*}
where we have used $\sum_{i=1}^{k-1}H_i = kH_{k-1} - (k-1).$ Combining $I_1$ and $I_2$ in \eqref{eqn:FisherFull} together with simplification, we obtain the Fisher information for the total observation of full times
\bsub\label{eq:Full_Fisher_Final}
\begin{equation}\label{eq:Full_Fisher_Final_a}
\mathbb{E}\left[\frac{\partial^2 \ell}{\partial a^2}  \right] = \frac{nk}{a^2}\Big( \big(\log(k+1)+ 1 +\alpha\big)^2 - 2 \big( H_{k-1}   -\gamma_e + \alpha \big)  -\frac{2}{k}\Big).
\end{equation}

If we consider an approximation for large $k$, we can apply $H_{k} \approx  \log k +\gamma_e + \bigoh(k^{-1})$ for $k\gg1$ and obtain the following compact representation
\begin{equation}\label{eq:Full_Fisher_Final_b}
\mathbb{E}\left[\frac{\partial^2\ell}{\partial a^2} \right]  \approx \frac{nk}{a^2}\Big( \big(\log k+\alpha\big)^2 +1 \Big) .
\end{equation}
\esub
The per-observation Fisher information is then given by
\begin{equation}
    I_{\text{full}} \approx \frac{k}{a^2}\Big( \big(\log k+\alpha\big)^2 +1 \Big).
\end{equation}
Hence we have derived two consistent estimators $\hat{a}$ based on either the difference in times \eqref{eq:dist_est} or the full arrival times \eqref{eq:rescaled_fullTimes}. The approximate convergence behavior for these estimates as $n\to\infty$ is
\begin{equation}\label{eq_final_fishers}
\sqrt{n}(\hat{a}_n - a) \to_\dd \mathcal{N}(0,I^{-1}), \qquad I(a) = \frac{1}{a^2} 
\left\{ \begin{array}{lc} k-1, & \mbox{time differences;} \\[4pt]
 k\, (\log k + \alpha)^2 + k, & \mbox{full times,} \end{array} 
\right.
\end{equation}
where $I(a)$ is the per-observation Fisher information.


\begin{figure}
\centering
\includegraphics[width=0.85\textwidth]{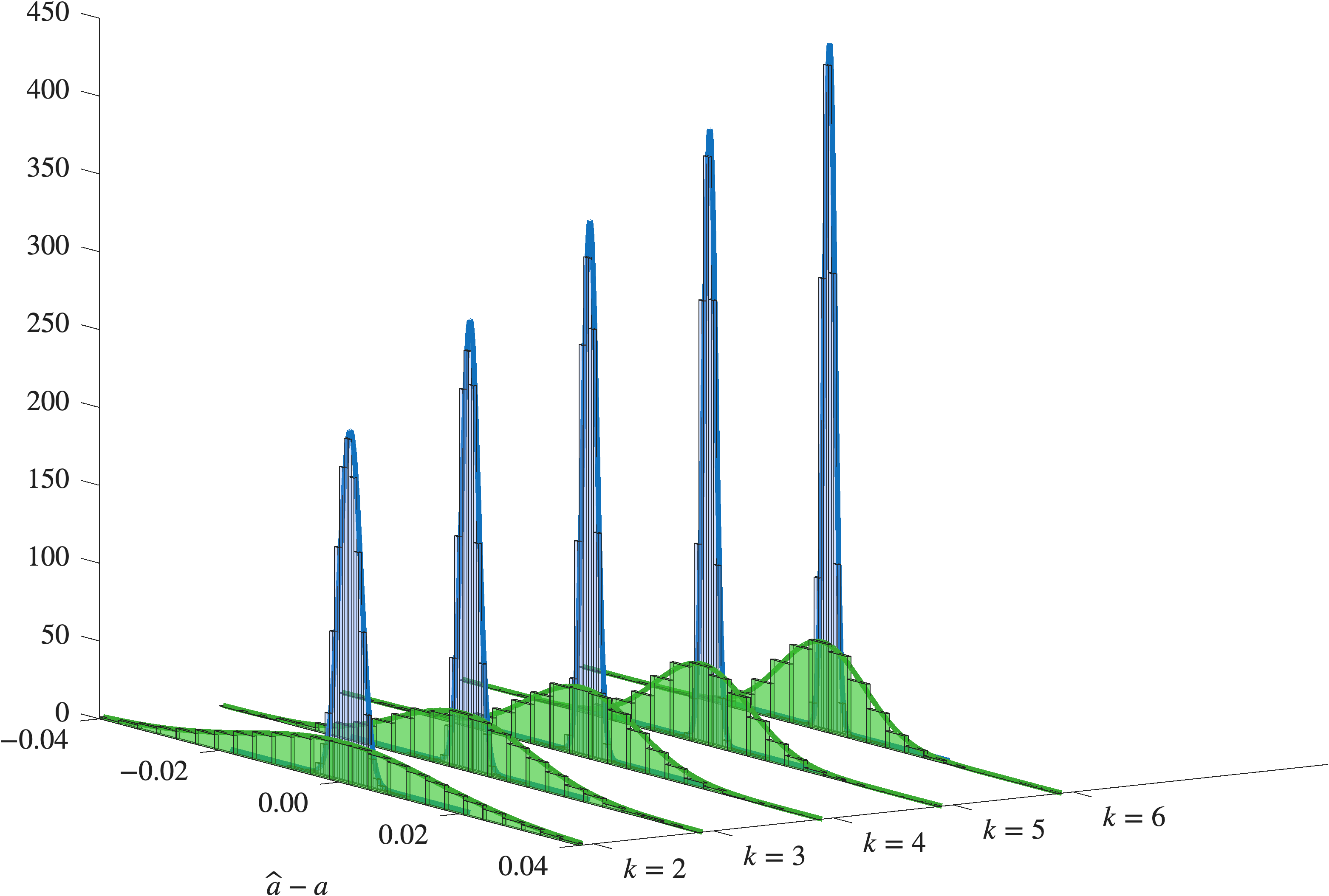}
\caption{Recovery of the extreme distribution parameter for test case $a=0.5$ for $n = 10^3$. Distribution in errors $(\hat{a}-a)$ using the full times (blue) and the time differences (green). Each histogram is formed by $M= 10^4$ repetitions and for each we overlay the distribution $\mathcal{N}(0,1/\sqrt{nI(a)})$ where the Fisher Information $I(a)$ is given in \eqref{eq_final_fishers}. \label{fig:FullTimePDF}}
\end{figure}

Before proceeding to numerical validation of the main result \eqref{eq_final_fishers}, we make some comments on its biological significance. Clearly $I_{\text{full}}>I_{\text{diff}}$ for any $k\geq1$ and hence knowledge of the full times yields more information on the source position. However, exploiting this fact requires storing these times and solving the nonlinear system \eqref{eq:rescaled_fullTimes}. Conversely, a cell can utilize a simpler procedure of a weighted average of the inter-arrival times \eqref{eq:dist_est}. While it exhibits higher variance (lower accuracy), this estimate requires substantially less computational work to obtain.

As a confirmation of the estimates \eqref{eq_final_fishers}, we take the fixed value of $a=1/2$ and generate $n = 10^3$ samples (from \eqref{eqn:dist_full}) of the first $k$ arrival times for $k=2,3,4,5,6$. From these observations, we create estimates $\hat{a}$ using both the MLE estimate based on the full times and the time differences. To explore the entire distribution predicted by \eqref{eq_final_fishers} we then repeat this process $M=10^4$ times and plot the results in Fig.~\ref{fig:FullTimePDF} \al{and observe that our result} \eqref{eq_final_fishers} accurately predicts the distribution of errors. In addition, these results confirm that the full time method generates estimates with much lower error and that the error is further reduced as $k$ increases.

\begin{figure}[htbp]
\centering
\subfigure[The distribution of errors $\hat{R}-R$ from full times (blue) and time differences (green) data for $N=10^6$ initial walkers of diffusivity $D=1$ and source $R=5$.]{\includegraphics[width=0.95\textwidth]{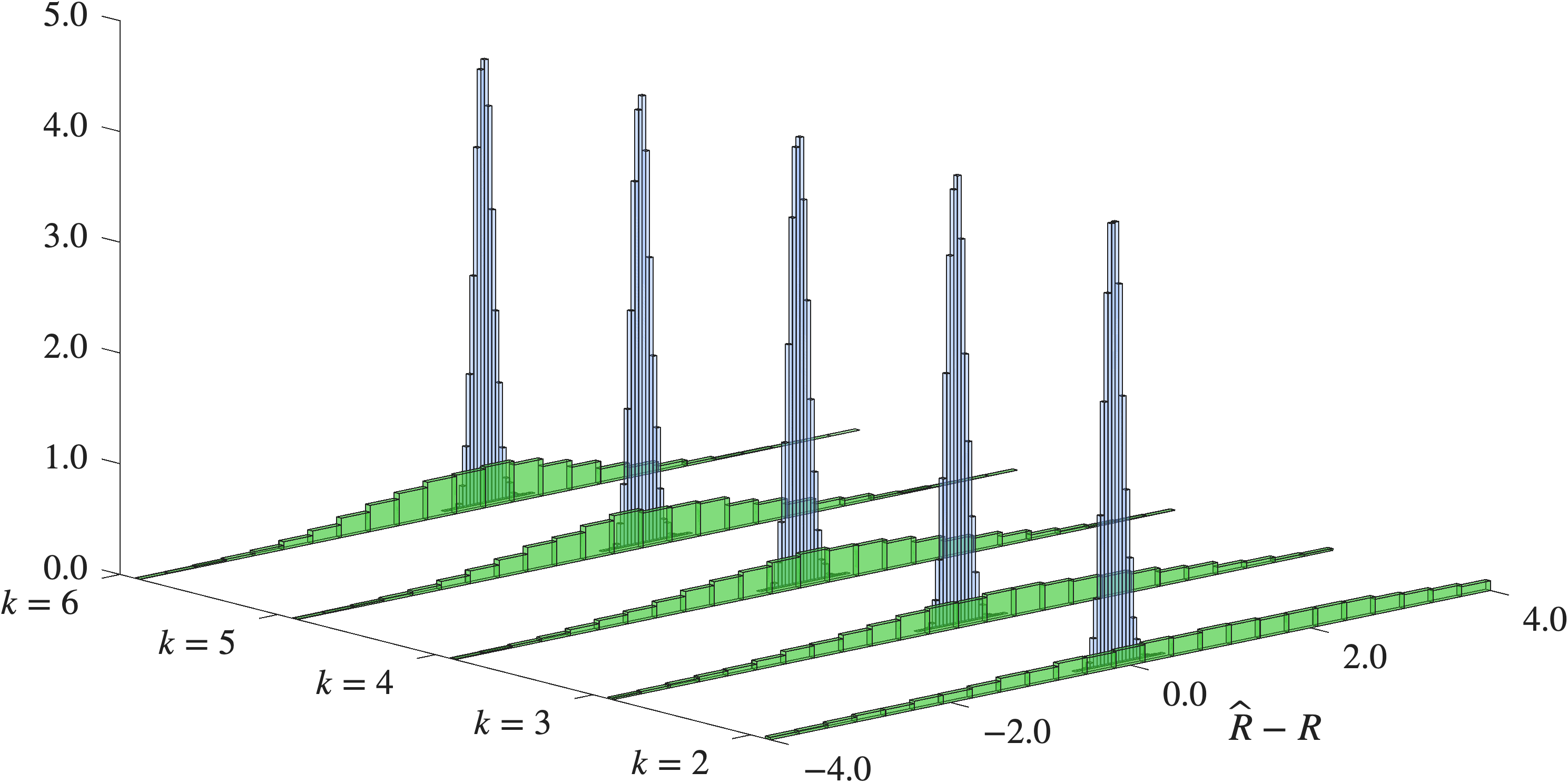}}\\[5pt]
\subfigure[Average error versus $k$ using the time differences.]{\includegraphics[width=0.445\textwidth]{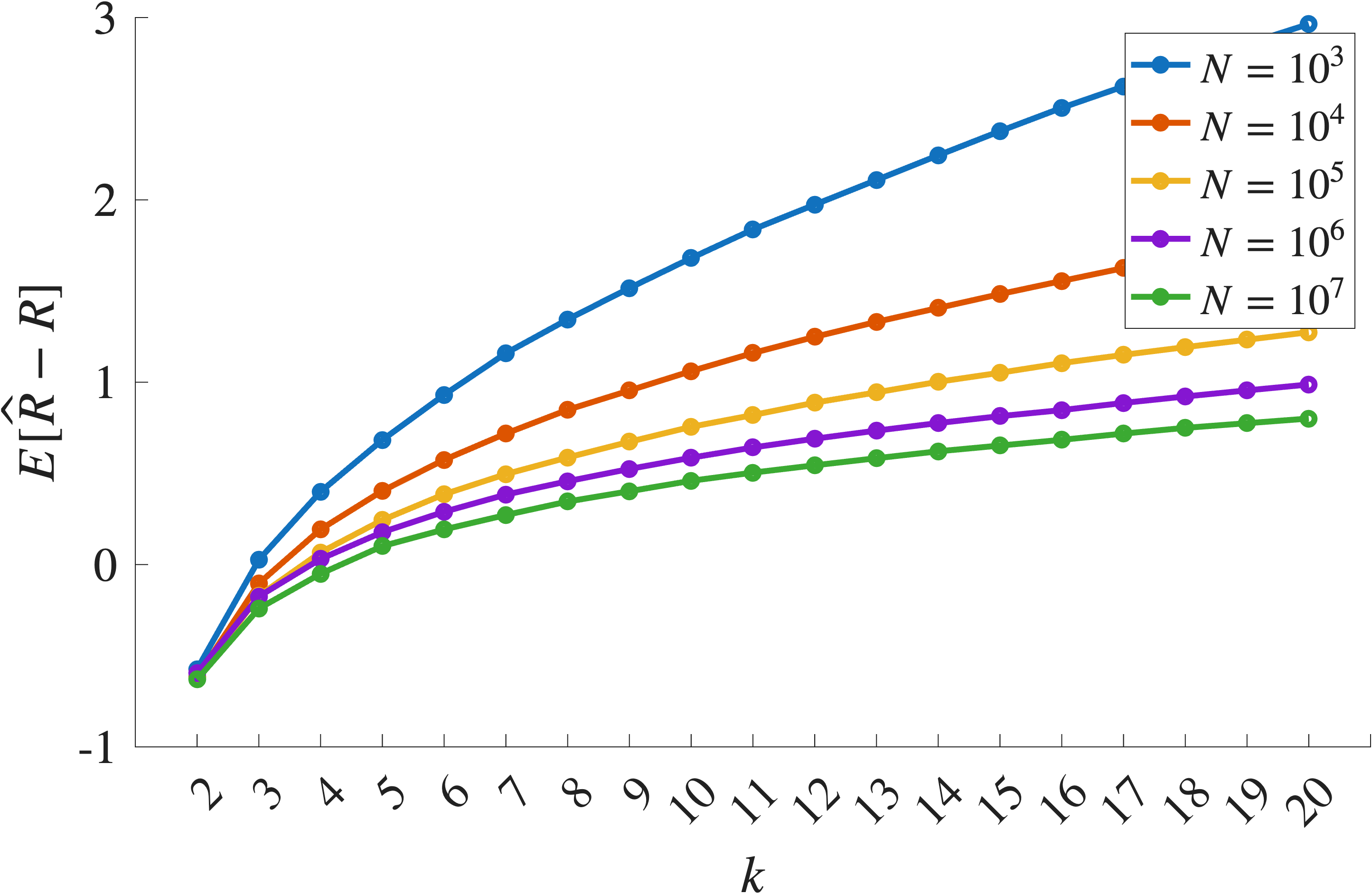}}
\qquad
\subfigure[Average error versus $k$ using the full set of times.]{\includegraphics[width=0.445\textwidth]{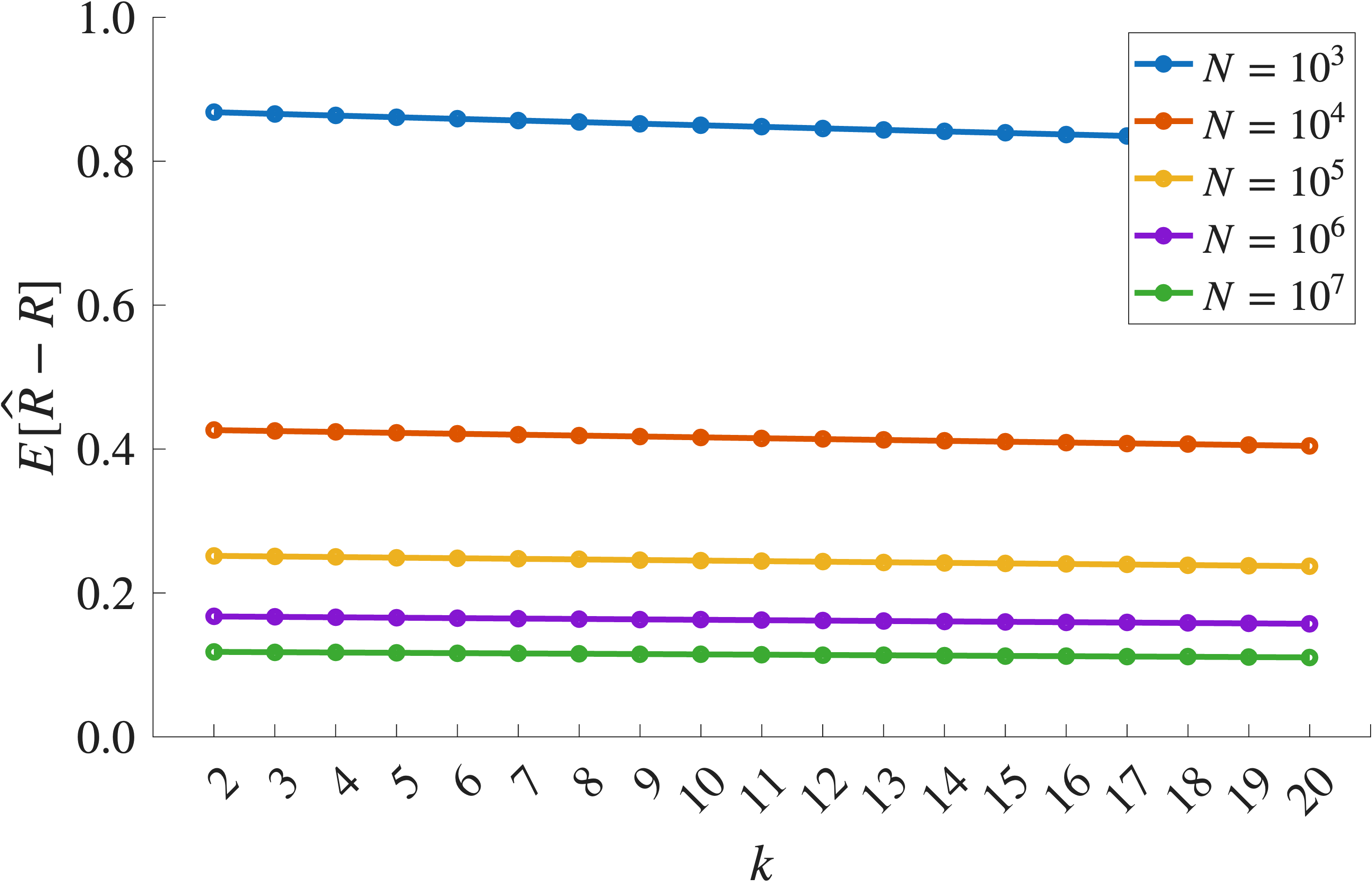}\label{fig:1DTimes_c}}
\caption{Accuracy in the prediction \eqref{eq:estimate_R} of source distance using the full arrival times and the differences in arrival times in one-dimension.  Panel (a): The distribution of errors for source detection when $R=5$ and $N = 10^6$. The estimate is vastly more accurate using the full times (blue) than the time differences (green). Distributions formed from $M=10^4$ independent estimates. In panels (b-c) we plot the average error against $k$ for initial walker numbers ranging from $N = 10^3-10^7$. 
\label{fig:1DTimes}}
\end{figure}

\subsection{Finite $N$}

The previous analysis and its demonstration were conducted purely on the limiting distribution as the number of walkers $N\to\infty$. To consider the problem for a finite number of walkers, we consider an idealized one-dimensional problem, where $N$ walkers with diffusivity $D$ originate at the point $x=R$ and are absorbed at $x=0$. The survival probability for this situation is given by $S(t) = 1 - \erfc(\frac{R}{\sqrt{4Dt}})$.
The limiting behavior as $t\to0^{+}$ is given by 
\[
1 - S(t) \sim \sqrt{\frac{4Dt}{\pi R^2}} e^{\frac{-R^2}{4Dt}}, \qquad \mbox{as} \qquad t\to0^{+}.
\]
By comparing with \eqref{eqn:main_Surv}, we identify the constants
\[
A = \sqrt{\frac{4D}{\pi R^2}}, \qquad p = \frac12, \qquad C = \frac{R^2}{4D}.
\]
Taking these values and applying them to the result \eqref{eqn:AandB}, we obtain that
\[
a_N = \frac{R^2}{DW(1+W)}, \qquad W= W_0\left(\frac{2N^2}{\pi}\right),
\]
where $W_0(z)$ is the principal branch of the LambertW function. We additionally notice that \al{$\alpha = \frac12(1+ W)$} is independent of $R$ and that $b_N= \alpha\, a_N$. Hence, given the first $k$ arrival times, we can form an estimate $\hat{a}$, either using the full times or their differences, to arrive at an estimate 
\begin{equation}\label{eq:estimate_R}
\hat{R} = [DW(1+W)\hat{a}]^{\frac12}.
\end{equation}
In Fig.~\ref{fig:1DTimes} we display results on the accuracy of the estimate $\hat{R}$ for a range of $k$ and initial walker numbers $N$ with the true value $R=5$. Across all examples, we see that exploiting the full arrival times yields a far more accurate estimate of the source distance $R$. \al{The observation that the full-time distribution is far narrower than the difference-based one is clear from the ratio $I_{\text{full}}/I_{\text{diff}} \approx (\log k + \alpha)^2 \approx 10^2$ for the values in Fig.~\ref{fig:1DTimes}. In Fig.~\ref{fig:1DTimes_c} we observe a very slight dependence on $k$ as the error is dominated by $N$ when the MLE estimator is so much more efficient.} The algorithms we have introduced and analyzed in the previous section, are based on the limiting distribution of the arrival times in the limit as $N\to\infty$. In Figs.~\ref{fig:1DTimes}(b-c) we observe the convergence of the error for a sequence of initial walker number $N$ and again notice that the full arrival time estimate performs markedly better than using only the differences.

\section{Estimation of source direction from extreme time data in 2D}\label{sec:2d_arrivals}

In this section, we move on to the problem of estimating the source direction from the information contained in the first $k$ arrival impacts. The arrival locations are time dependent since quick arrivals are more focused around the source direction whereas longer sojourns arrive more uniformly on the cell surface.

\subsection{Statistics of arrival locations for a circular cell.}

In order to determine the impact location distribution, we consider a circular cell of unit radius and a source placed at $\bx_0 = (R,0)$ for $R>1$. The placement of the source at an angle $\theta_0$ relative to the horizontal is a simple rotation in the counter-clockwise direction. \al{In polar coordinates $\bx = r(\cos\theta,\sin\theta)$, the quantity $u(r,\theta,t)r\mathrm{d}r\mathrm{d}\theta$ is the probability of a particle being in $(r,r+\mathrm{d}r) \times (\theta,\theta+ \mathrm{d}\theta)$ at time $t$. Then  $u(r,\theta,t)$ solves the diffusion equation}
\bsub\label{eq:prob_full}
\begin{gather}
\frac{\partial u}{\partial t} = D\Big [\frac{\partial^2 u}{\partial r^2}  + \frac{1}{r} \frac{\partial u}{\partial r} +  \frac{1}{r^2} \frac{\partial^2 u}{\partial \theta^2}\Big], \quad r\in(1,\infty), \quad \theta \in (-\pi,\pi) \quad t>0;\\[5pt]
u = \frac{1}{r}\delta(r-R)\delta(\theta), \quad r\in(1,\infty), \quad \theta \in (-\pi,\pi)\quad t=0;\\[5pt]
u(r,\theta+2\pi,t) = u(r,\theta,t), \quad r\in(1,\infty), \quad \theta \in(-\pi,\pi), \quad t \geq 0;\\[5pt]
\al{\partial_{\theta}u(r,\theta+2\pi,t) = \partial_{\theta}u(r,\theta,t), \quad r\in(1,\infty), \quad \theta \in(-\pi,\pi), \quad t \geq 0.}
\end{gather}
The choice of boundary conditions on the surface of the cell can be influenced by many factors including the distribution and affinities of the receptors. For the purposes of the present study, we consider a continuum of receptors evenly distributed on the cellular surface. Under these conditions, and assuming $N_{\text{rec}}$ \al{receptors of non-dimensional common radius $\eps_{\text{rec}}$,} the following Robin boundary condition has recently \cite{LindsayRSOS2023} been derived
\begin{equation}\label{eq:robin}
D\frac{\partial u}{\partial r} = \kappa u, \quad r = 1; \qquad D\kappa^{-1} = -\frac{2}{N_{\text{rec}}} \log \Big(\frac{N_{\text{rec}}\eps_{\text{rec}}}{4}\Big).
\end{equation}
\esub
When the number of receptors is large, $(N_{\text{rec}}\approx 10^5)$, each with radius $10^{-8}-10^{-10}m$ \cite{GPCR_size}, the value of $\kappa$ is large, resulting in an effectively fully absorbing boundary. In the limiting $\kappa\to\infty$ case, the Dirichlet boundary condition $u=0$ can be applied on $r=1$. Further analysis on the homogenization of porous membranes is described in \cite{Brennan:2026,lindsay2015}.

The treatment of receptors as a uniform distribution rather than considering their discrete positions (e.g.~\cite{LindsayRSOS2023,LLM2020}) affords many simplifications. \al{It is frequently observed that receptors can have non-uniform surface distributions that are modulated by the curvature of the membrane  \cite{Endres2025}. Hence, the analysis of the uniform case presented here serves as a baseline for future studies on the role of geometry in chemotaxis.}

\al{With the assumption of the uniform boundary condition \eqref{eq:robin}}, we are able to determine the short time asymptotics of the solution to \eqref{eq:prob_full}. The two key quantities that describe the hitting times and hitting locations, are the survival probability $S(t)$ and surface flux $\J(\theta,t)$, respectively. These quantities are defined from the solution of \eqref{eq:prob_full} by
\bsub\label{eq:FluxandS}
 \begin{align}
 S(t) &= \int_{r=1}^{\infty} \int_{\theta=-\pi}^{\pi}u(r,\theta,t) \, r\, \dd r\, \dd\theta \\[5pt]
 \J(\theta,t) &=D\partial_ru(r=1,\theta,t).
\end{align}
\esub
To derive the short time asymptotic forms of \eqref{eq:FluxandS} as $t\to0$, we apply the method of moments \cite{LindsayRSOS2023}. The first step is to apply the Laplace transform $\hat{u}(r,\theta;s)=\int_{t=0}^{\infty} u(r,\theta,t)e^{-st} \dd t$ to equation \eqref{eq:prob_full}. We follow a standard process of writing the solution in terms of the cosine-Bessel series \cite{moth2020} where we enforce continuity and a jump condition at $r=R$, require the solution to be finite as $r\to\infty$ and also enforce the boundary condition \eqref{eq:robin}. The Laplace transform of the flux is found to be
\begin{equation}\label{eqn:LT_flux}
\hat{\J}(\theta;s) = \frac{1}{2\pi} \chi_0(q) + \frac{1}{\pi}\sum_{n=1}^{\infty} \chi_n(q) \cos n\theta, \qquad \chi_n(q) = \frac{\kappa \big(K_n(q R)/K_n'(q)\big) }{\kappa \big(K_n(q)/K'_n(q)\big) - D q}.
\end{equation}
where $q = \sqrt{s/D}$. Our goal is to approximate the solution of \eqref{eqn:LT_flux} for $q\to\infty$ which corresponds to the limit $t\to0^{+}$ in \al{the time domain}. We define the moments $\widehat{M}_n(s) = \int_{-\pi}^{\pi}\hat{\J}(\theta;s)\theta^n \dd\theta$ and apply $\cos n\theta \approx 1 - \frac{1}{2}n^2\theta^2 + \frac{1}{24} n^4 \theta^4$ in \eqref{eqn:LT_flux} to obtain
\begin{equation}\label{eqn:MoM_main}
\chi_n(q) = \int_{-\pi}^{\pi} \hat{\J}(\theta;s) \cos n \theta \, \dd \theta \approx \widehat{M}_0(s) - \frac{n^2}{2} \widehat{M}_2(s) + \frac{n^4}{24} \widehat{M}_4(s).
\end{equation}
The large argument asymptotics \cite[\S9.7.2]{abramowitzstegun} of $K_n(z)$ for $z\to\infty$ yields the behavior
\begin{align*}
\frac{K_n(q R)}{K_n'(q)} \sim & \frac{e^{-q(R-1)}}{R^{\frac12}} \left[ -1 + \frac{ (4n^2+3) (R-1) +4}{8Rq}  \  + \right. \\
& \left. \frac{(-16n^4 + 8n^2 -33)(R-1)^2 + (32n^2-72)(R-1) + (64n^2 -48)}{128 R^2q^{2}} \right].
\end{align*}
Notice that the term $K_n(q)/K'_n(q)$ arises from substituting $R=1$ into this expression. We insert this approximation into the expression for $\chi_n$ given in \eqref{eqn:LT_flux} and then apply to \eqref{eqn:MoM_main}. Gathering terms at the relevant orders of $n$, we obtain the following expressions for the moments
\[
\widehat{M}_0 \sim \frac{\kappa e^{-q(R-1)}}{R^{\frac12}(\kappa + D q)}, \quad \widehat{M}_2 \sim \frac{\kappa(R-1) e^{-q(R-1)}}{R^{\frac32}q(\kappa + D q)}, \quad \widehat{M}_4 \sim \frac{3\kappa(R-1)^2 e^{-q(R-1)}}{R^{\frac52}q^2(\kappa + D q)}, 
\]
in the limit as $q \to \infty$. Inverting these Laplace transforms explicitly as $q\to\infty$ ($t\to0^{+}$), we identify the following limits for the variance and kurtosis of the flux
\[
\text{Var}[\J(\theta,t)] := \frac{M_2}{M_0} \sim \frac{2 Dt}{R}, \qquad \text{Kur}[\J(\theta,t)] := \frac{M_4 M_0}{[M_2]^2} \sim 3, \qquad t \to 0^{+}.
\]
\al{The kurtosis is the normalized fourth moment $\text{Kur}[\mathcal{J}] = M_4M_0/M_2^2$, equal to $3$ for a Gaussian; its convergence to $3$ as $t\to0^{+}$ confirms the limiting flux is Gaussian.} Hence, in the limit as $t\to0^{+}$, we have derived from \eqref{eq:prob_full} the short time behavior
\bsub\label{eqn:short_time_asy}
\begin{align}\label{eqn:short_time_asy_a}
\J(\theta,t) \sim \frac{M_0(t)}{\sigma \sqrt{2\pi}} e^{-\frac{\theta^2}{2\sigma^2}}, \qquad \sigma^2 \sim \frac{2Dt}{R},
\end{align}
as $t\to 0^{+}$. Here the total arrival rate $M_0(t)$ at the cell surface is given by
\begin{equation}\label{eqn:short_time_asy_b}
M_0(t) \sim  \left\{ 
\begin{array}{ll}
\frac{\kappa(R-1)}{\sqrt{\pi R Dt}} e^{-\frac{(R-1)^2}{4Dt}}\left[ \frac{1}{2\kappa t + (R-1) }\right], & \qquad \kappa\in(0,\infty);\\[5pt]
\frac{(R-1)}{\sqrt{4\pi R Dt^3}} e^{-\frac{(R-1)^2}{4Dt}} & \qquad \kappa = \infty,
\end{array}
\right.
\end{equation}
as $t\to0$. The survival probability satisfies $S'(t) = -\int_{\theta=-\pi}^{\pi} \J (\theta,t) \dd\theta \sim -M_0(t)$. From this relationship, we obtain that
\begin{equation}\label{eqn:short_time_asy_c}
S(t) \sim 1 - \left\{ 
\begin{array}{ll}
\frac{4\kappa}{ \sqrt{\pi R} }  \frac{\sqrt{D} }{ (R-1)^2 } t^{\frac32} e^{-\frac{(R-1)^2}{4Dt}}, & \qquad \kappa\in(0,\infty);\\[5pt]
\frac{ \sqrt{4D} }{ (R-1)\sqrt{\pi R} } t^{\frac12} e^{-\frac{(R-1)^2}{4Dt}}, & \qquad \kappa = \infty.
\end{array}
\right.
\end{equation}
\esub
Hence, comparing between \eqref{eqn:short_time_asy_c} and \eqref{eqn:main_Surv}, the constants from the short time survival probability $S(t) \sim 1- At^p e^{-\frac{C}{t}}$ that characterize the distribution of arrival times are
\begin{equation}\label{eq:constants}
\begin{array}{llll}
p = \frac{3}{2}, & \quad A = \frac{4\kappa}{ \sqrt{\pi R} }  \frac{\sqrt{D} }{ (R-1)^2 }, & \quad C = \frac{(R-1)^2}{4D},    & \qquad \kappa \in (0,\infty);\\[5pt]
p = \frac{1}{2}, & \quad A = \frac{ \sqrt{4D} }{ (R-1)\sqrt{\pi R} },                              & \quad C = \frac{ (R-1)^2 }{ 4D }, &\qquad \kappa = \infty.
\end{array}
\end{equation}
These variables now define the scalars $a_N$ and $b_N$ that characterize the distribution of arrival times through the definitions \eqref{eqn:AandB}.

\subsection{The angular distribution of impacts}

\begin{figure}[htbp]
\centering
\subfigure[Distribution of $\theta_{k,N}$ for a range of $R$ values.]{\includegraphics[width = 0.55\textwidth]{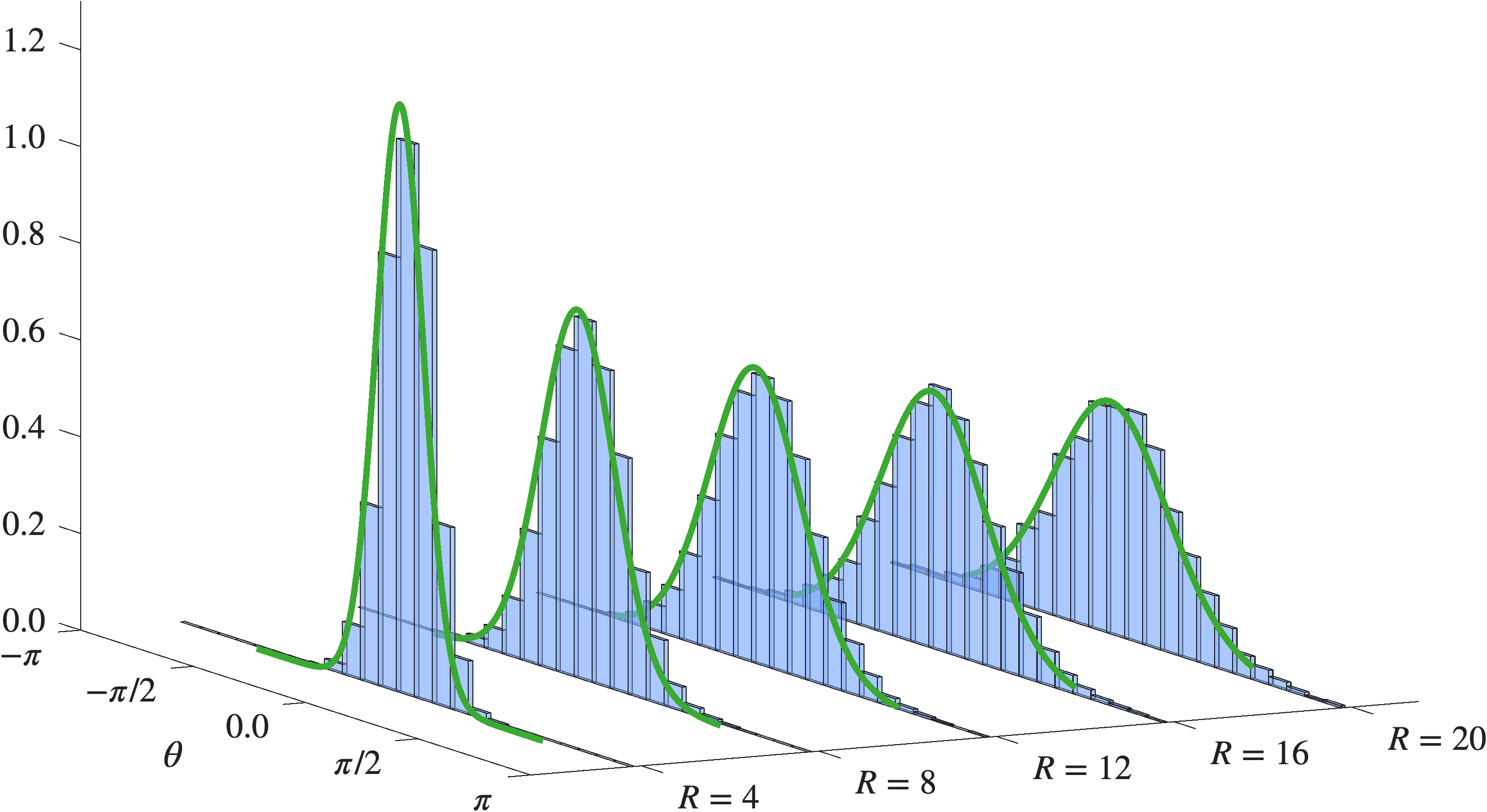}\label{fig:FluxSurface_a}}
\subfigure[Distribution of $\theta_{k,N}$ for a range of $k$.]{\includegraphics[width = 0.375\textwidth]{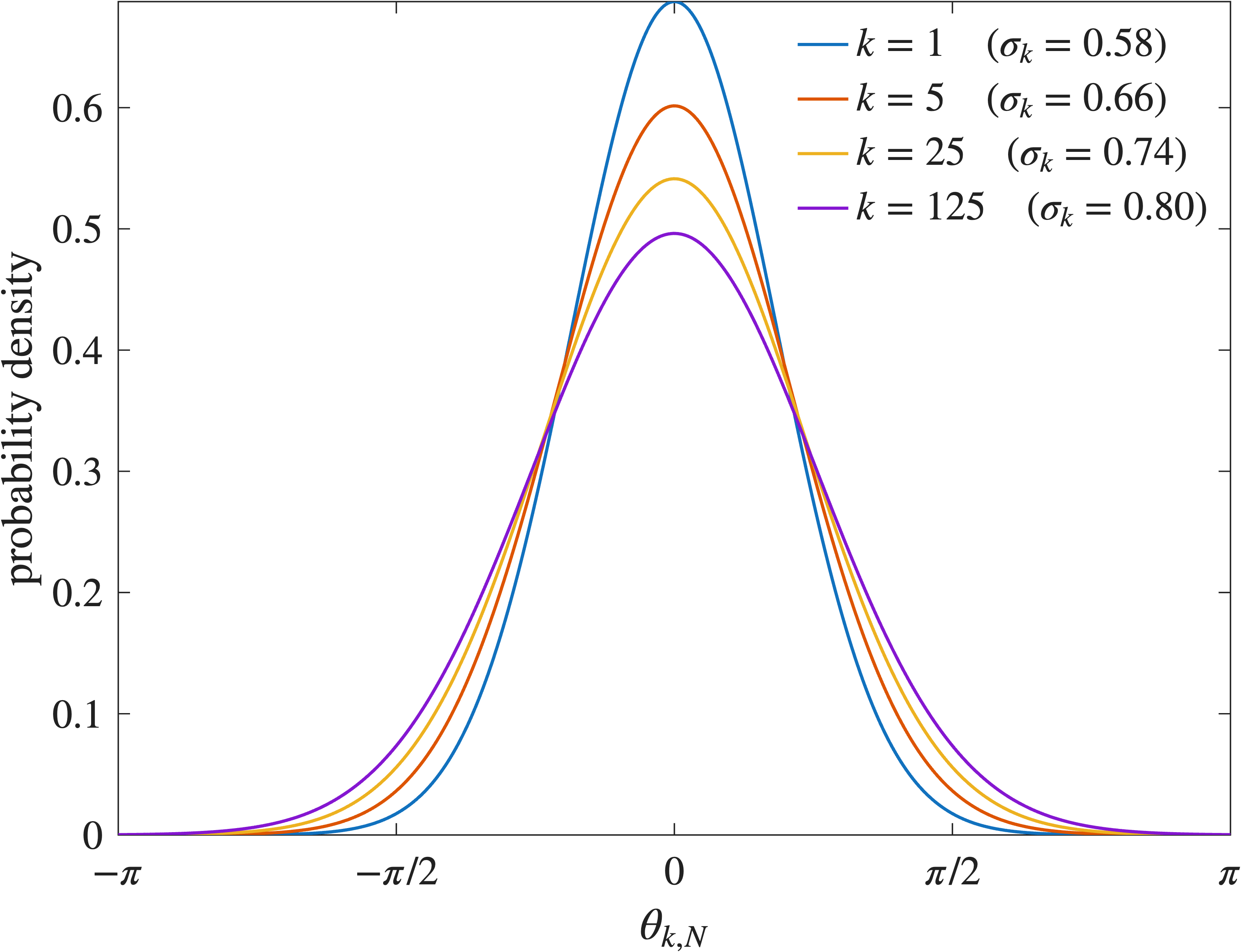}\label{fig:FluxSurface_b}}
\caption{The distribution of impact angles $\theta_{k,N}$. Panel (a) shows results for $k=10$, $D=1$, $\theta_0 = 0$ and $N= 10^6$ for various $R$. The histograms are based on $M=10^4$ impacts sampled directly from the case of Dirichlet boundary condition ($\kappa = \infty$). The solid curve is the predicted limiting distribution \eqref{eqn:AngleDist} with good agreement observed. \al{In panel (b) we show the limiting impact distribution \eqref{eqn:AngleDist} with fixed parameters $R= 5$, $D=1$, $N=10^3$ and over a range of $k$ values. The distribution of arrival angles broadens as $k$ increases.} \label{fig:FluxSurface} }
\end{figure}

Here we analyze the $k^{\mathrm{th}}$ arrival location characterized by the impact time $T_{k,N}$ and determine the distribution of the corresponding angular location $\theta_{k,N}$. As discussed previously, the distribution of the angular location depends on the time of impact. Hence, we have that
\[
\mathbb{P}[\theta_{k,N} = \Theta] = \int_{t=0}^\infty \mathbb{P}[\theta_{k,N} = \Theta | T_{k,N}=t]  \mathbb{P}[T_{k,N} = t] \dd t.
\]
\al{The limiting distribution for $\mathbb{P}[T_{k,N}= t]$ is given in \eqref{eqn:T_dist} and takes the form
\[
\mathbb{P}[T_{k,N}= t] = \frac{1}{a_N(k-1)!}\exp[\varphi(t)], \qquad \varphi(t) = \frac{k(t-b_N)}{a_N} - \exp\left[\frac{t-b_N}{a_N}\right].
\]
The distribution of angular arrivals now satisfies 
\begin{equation*}
\mathbb{P}[\theta_{k,N} = \Theta] = \frac{1}{a_N(k-1)!}  \int_{t=0}^\infty \mathbb{P}[\theta_{k,N} = \Theta | T_{k,N}=t] \exp[\varphi(t)]\dd t.
\end{equation*}
In the limit as $N\to\infty$, which corresponds to $a_N\to0$, we apply Laplace's method to simplify this integral. The value of the integral is dominated by the sharp peak of the function $\varphi(t)$ around the peak (mode) $t_c =b_N + a_N \log k$. In the vicinity of $t_c$, the function has a local quadratic shape with $\varphi''(t_c)= -k/a_N^2$. This allows for a standard application of Laplace's method.} From the distribution of fluxes along the surface of the cell derived in \eqref{eqn:short_time_asy_a}, we conclude that
\begin{equation}\label{eqn:sig2}
\mathbb{P}[\theta_{k,N} = \Theta \ | \ T_{k,N}= t_c ] = \frac{1}{\sigma \sqrt{2\pi}} e^{-\frac{\theta^2}{2\sigma^2}}, \qquad \sigma^2 \sim \frac{2Dt_c}{R}.
\end{equation}
Finally, we conclude that the impact locations have a Gaussian distribution given by
\begin{equation}\label{eqn:AngleDist}
\theta_{k,N} \sim \mathcal{N}\big(\theta_0, \sigma_{k,N}^2\big), \qquad \sigma_{k,N}^2 = \frac{2D}{R}\Big(b_N + a_N \log k \Big),
\end{equation}
in the limit as $N\to\infty$.  From equation \eqref{eqn:AngleDist} we observe that since $a_N>0$, particles arriving earlier to the cell have impact locations more tightly distributed (lower variance) around the source direction. Consequently, the first particle to arrive at the cell has the most directional information, followed by the second, and then followed by the third, etc. As a confirmation of this, we plot in Fig.~\ref{fig:FluxSurface_a} a comparison between particle simulations of the arrival distribution and the result \eqref{eqn:AngleDist}. \al{In Fig.~\ref{fig:FluxSurface_b} we plot the limiting distribution of the $k^{\mathrm{th}}$ impact \eqref{eqn:AngleDist} for fixed $R$ and a range of $k$ values.}

\al{The variance in the impact location \eqref{eqn:AngleDist} is independent of the particle diffusivity $D$. This reflects a cancellation of two competing effects: the angular spread of the flux conditioned on arrival at time $t$ grows as $2Dt_c/R$ while the characteristic arrival time of the $k^{\mathrm{th}}$ molecule, $t_c = b_N + a_N \log k$, scales as $1/D$. A larger diffusivity therefore delivers the $k^{\mathrm{th}}$ impact proportionally sooner by exactly the factor by which it broadens the angular distribution. The result is that the $k^{\mathrm{th}}$ arrival always samples the same range of angular space hence $D$ sets the timescale of capture but not the geometry of the early-impact distribution.}

\section{Recovery of source direction from extreme arrivals}\label{sec:Source_Extreme}

In this section, we use the main result \eqref{eqn:AngleDist} to explore the accuracy of several estimators on the direction of the source.

\subsection{Assuming the cell accesses and averages the full angle values} Let us consider a situation where the cell uses the first $k$ angles $\{\theta_{j,N}\}_{j=1}^k$ from $N$ initial walkers to form an estimate \al{of the source direction. Without loss of generality, we set the source to have direction $\theta_0=0$.} A first quantity to consider is the average angle $\Theta^{\text{avg}}_k$ inferred by these impacts. It has the following definition:
\begin{equation}\label{eqn:theta_avg}
\Theta^{\text{avg}}_k = \frac{1}{k}\sum_{j=1}^k \theta_{j,N}, \qquad [\text{Average angle}].
\end{equation}
Applying the distribution of each $\theta_{j,N}$ for $j = 1,\ldots,k$ given in \eqref{eqn:AngleDist}, we obtain that
\begin{equation}\label{eqn:AvgAngle} 
\Theta^{\text{avg}}_k \sim \mathcal{N}\left(0, \frac{2D}{Rk^2}\Big(b_Nk + a_N\log k! \Big)\right) \approx \mathcal{N}\left(0, \frac{2D}{Rk}\Big(b_N -a_N + a_N\log k\Big) \right),
\end{equation}
where on the last step we have applied the approximation $\log k! \approx k\log k - k $. We notice that by increasing the number $k$ of points averaged, the variance decreases like $k^{-1}\log k$. This indicates that although the variance associated with individual impacts increases with $k$, the averaged variance is overall decreasing.

\subsection{Estimates based on the difference in angles} Here we consider \al{an} estimate based on the \al{average of the differences in} the angle between impacts. \al{Specifically, we derive the distribution of
\begin{equation}
\Theta_k^{\text{diff}} = \frac{1}{k}\sum_{j=1}^{k-1} \big[ \theta_{j+1,N} - \theta_{j,N} \big] 
\end{equation}
} From \eqref{eqn:AngleDist}, we have that $\theta_{k,N}\sim\mathcal{N}(0,\sigma_{k,N}^2)$, and hence we observe that
\begin{equation}
\theta_{j+1,N} - \theta_{j,N} \sim \mathcal{N}\left(0, \frac{2D}{R}\Big(b_N + a_N\log\big(j+1\big)\Big) + \frac{2D}{R}\Big(b_N + a_N \log\big(j\big)\Big)\right).
\end{equation}

By summing the first $k-1$ differences in angle we get the following simplification
\begin{align*}
\sum_{j=1}^{k-1} \big[\theta_{j+1,N} - \theta_{j,N} \big] &= (\theta_{2,N} - \theta_{1,N}) + (\theta_{3,N} - \theta_{2,N}) + \cdots + (\theta_{k,N} - \theta_{k-1,N})\\
& =  \theta_{k,N} - \theta_{1,N}.
\end{align*}
The average of the differences then has the distribution
\begin{equation}\label{eq:EqnDiff}
\Theta_k^{\text{diff}} = \frac{1}{k}\sum_{j=1}^{k-1} \big[ \theta_{j+1,N} - \theta_{j,N} \big] = \frac{1}{k} (\theta_{k,N}-\theta_{1,N})\sim \mathcal{N}\left(0, \frac{2D}{Rk^2}\big(2b_N + a_N\log k\big)\right).
\end{equation}

\begin{figure}
\centering
\includegraphics[width=0.9\textwidth]{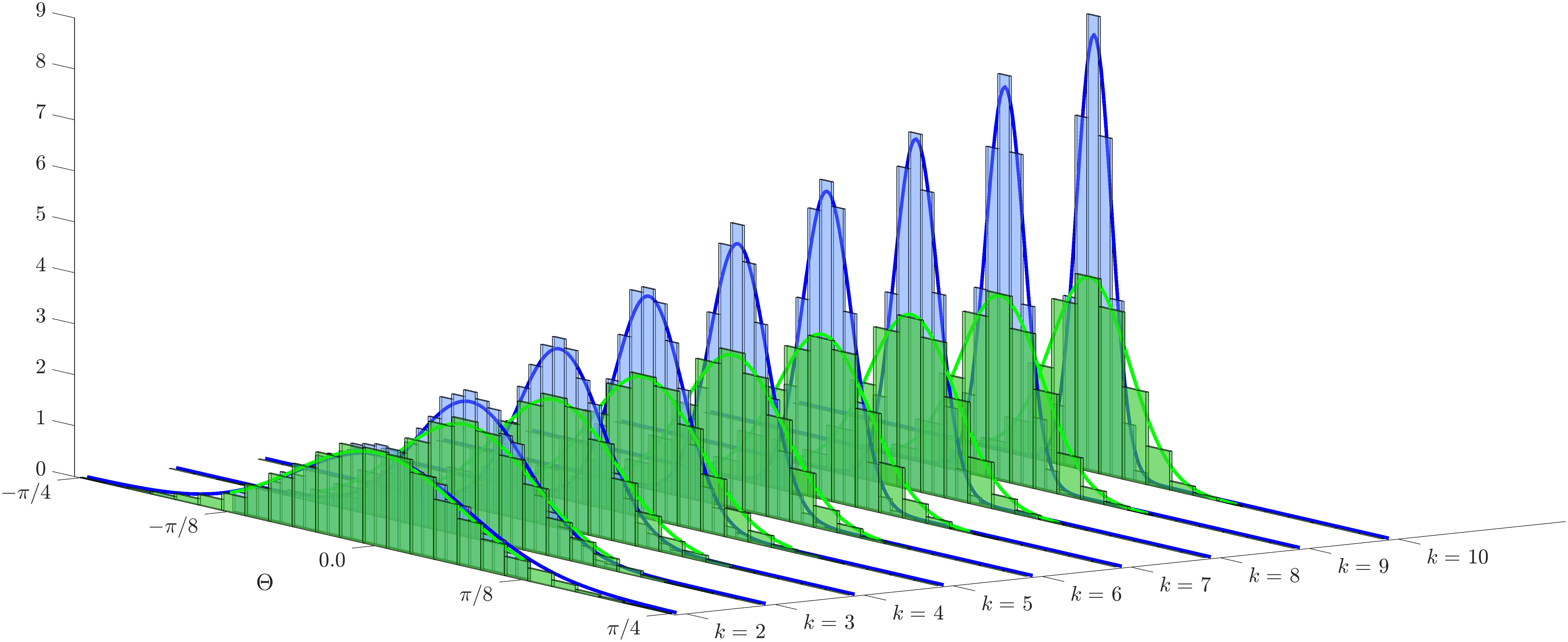}
\caption{Distribution of angular estimates in predictions using the average angles $\Theta_k^{\text{avg}}$ (green histograms) defined in \eqref{eqn:AvgAngle} and the differences estimate $\Theta_k^{\text{diff}}$ (blue histograms) defined in \eqref{eq:EqnDiff}. \label{fig:AvgVsDiffDifferences}}
\end{figure}

In Fig.~\ref{fig:AvgVsDiffDifferences} we show confirmation of the expressions \eqref{eqn:AvgAngle} for the average $\Theta_k^{\text{avg}}$ and \eqref{eq:EqnDiff} for the average difference $\Theta_k^{\text{diff}}$. We generated $M=10^4$ samples of the $k^{\mathrm{th}}$ arrival angle $\theta_{k,N}$ to a circular cell based on $N=10^6$ initial walkers. These were sampled exactly from the closed form flux expression \eqref{eqn:LT_flux} by applying a numerical inverse Laplace transform (see \cite{CHERRY2025}). We observe that for $k=1$, both estimates provide similar accuracy on the source direction. However, as $k$ increases the difference estimate \eqref{eq:EqnDiff} improves because fewer values are averaged.

\subsection{The statistics of the resultant vector} In this example, we analyze the resultant formed by averaging over a vector of each arrival. Specifically, let 
\[
\textbf{n}^{\text{res}} \equiv \frac{1}{k} \sum_{j=1}^k \textbf{n}_j, \qquad \textbf{n}_j = \big(\cos\theta_j, \sin\theta_j\big), \qquad j = 1,\ldots,k,
\]
be the equally weighted average of the first $k$ impacts  at the cell. When utilizing a polar coordinate system where the source has location $\bx_0 = R(1,0)$, the true source is in the direction $\textbf{e}_1 = (1,0)$. Hence, the angular difference between the vector $\textbf{n}^{\text{res}}$ and the source is estimated by
\begin{equation}\label{eq:Res_Est}
\Theta^{\text{res}} = \braket{ \textbf{n}^{\text{res}},\textbf{e}_1 }  = \frac{1}{k} \sum_{j=1}^k \cos \theta_j.
\end{equation}
In the limit as $N\to\infty$, we assume the angular distribution is tightly distributed around the region $|\theta|\ll1$. Expanding for $|\theta_j|\ll1$, we have that
\begin{equation}\label{eq:ChiSqrd}
\Theta^{\text{res}} =   \frac{1}{k} \sum_{j=1}^{k} \cos \theta_j \approx 1 - \frac{1}{2k}\sum_{j=1}^k \theta_j^2. 
\end{equation}
\al{The expression \eqref{eq:ChiSqrd}, which represents a combination of squared Gaussian random variables, is known to follow the \emph{generalized chi-squared} distribution. In the simpler scenario in which $\theta_j$ are i.i.d., the random variable follows the chi-squared distribution.} 

\begin{figure}
\centering
\includegraphics[width = 0.75\textwidth]{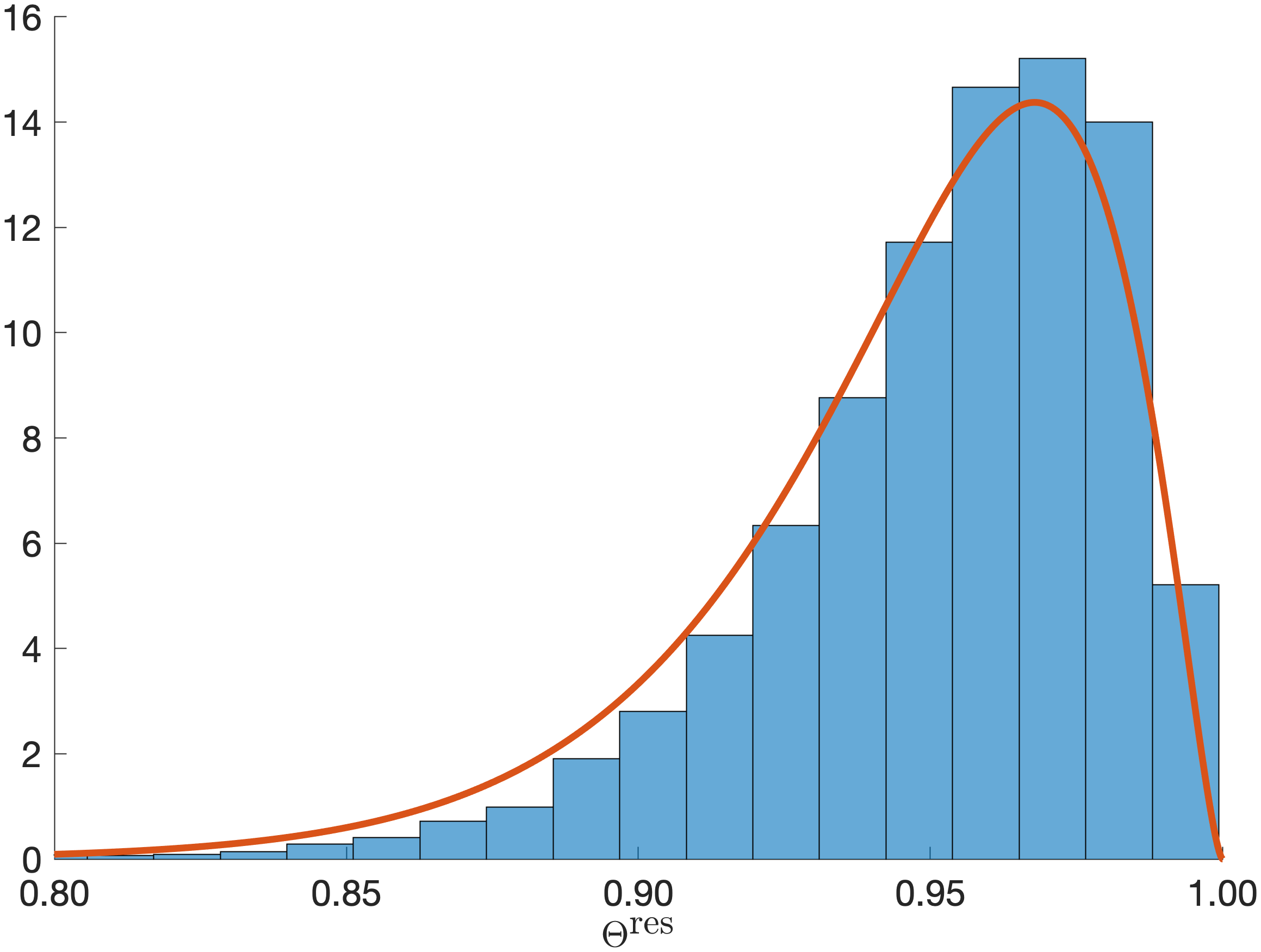}
\caption{The distribution of the resultant vector based on the average of the first $k$ arrivals for $k= 5$ and $N= 10^6$ initial walkers at a distance of $R=4$. The histogram shows estimates $\Theta^{\text{res}} = {\textbf n}^{\text{res}} \cdot \textbf{e}_1$, defined in \eqref{eq:Res_Est} for $M=10^4$ realizations. The generalized chi-squared distribution (solid) defined in \eqref{eq:ChiSqrd} gives a close approximation to the estimate. \label{fig:genchi2} }
\end{figure}

Hence the error $\rho_k^{\text{res}}$ in the resultant vector formed by $k$ arrivals is described by the generalized chi-squared variable \cite{genchi2} 
\begin{equation}\label{eq:error_resultant}
\rho_k^{\text{res}} = \frac{1}{2k}\sum_{j=1}^k \theta_j^2 = \frac{D}{Rk} \sum_{j=1}^{k} \big(b_N+a_N\log(j)\big)Z_j^2,
\end{equation}
where $Z_j\sim \mathcal{N}(0,1)$. The expected value and variance of this random variable satisfy
\bsub\label{eqn:MeanVar}
\begin{align}
\label{eqn:MeanVar_a} \mathbb{E}[\rho_k^{\text{res}}] &=  \frac{D}{Rk} \sum_{j=1}^{k} \big(b_N+a_N\log j\big) \approx \frac{D}{R}\Big(b_N + a_N \big(\log k -1\big)\Big), \\[4pt]
\label{eqn:MeanVar_b} \text{Var}[\rho_k^{\text{res}}] &= \frac{2D^2}{R^2k^2} \sum_{j=1}^{k} \big(b_N+a_N\log j\big)^2 \approx \frac{2D^2}{R^2k}\Big( (a_N \log k + b_N -a_N )^2 + a_N^2 \Big).
\end{align}
\esub
In arriving at \eqref{eqn:MeanVar}, we have applied the approximations 
\[
\sum_{j=1}^{k} \log j \approx k\log k - k, \qquad\sum_{j=1}^{k} [\log j]^2 \approx k\,\big(\log^2 k  - 2\log k + 2\big), \qquad k\gg1.
\]
In Fig.~\ref{fig:genchi2}, we plot results for the estimate $\Theta^{\text{res}}$ for $R=4$ and $k=5$. We see that the distribution of the resultant is closely approximated by the generalized chi-squared distribution \eqref{eq:error_resultant}. This good agreement validates the predicted mean and variance of the estimator.

\al{\paragraph{Bias-variance trade-off for finite $k$}
The observation that the error \eqref{eqn:MeanVar_a} increases gradually with $k$, while the variance \eqref{eqn:MeanVar_b} decreases suggests a classic bias-variance trade-off where a small number of arrivals can combine in a simple vector average to give an accurate estimate of the source direction. To investigate this further, we calculate the mean squared error (MSE), to be
\begin{align}
\nonumber\mathbb{E}\big[(\rho_k^{\text{res}})^2\big] &= \mathbb{E}\big[\rho_k^{\text{res}}\big]^2 + \text{Var}\big[\rho_k^{\text{res}}\big]\\[5pt]
    \nonumber & = \frac{D^2}{R^2}\left[ \Big(b_N + a_N \big(\log k -1\big)\Big)^2  + \frac{2}{k} \Big(  \big(a_N(\log k-1) + b_N\big)^2 + a_N^2  \Big)\right]\\
    & = \frac{D^2a_N^2}{R^2} \mu(k), \qquad \Big[\, \mu := \Big(\alpha + \log k - 1\Big)^2 \Big(1+ \frac{2}{k}\Big) + \frac{2}{k} \, \Big], \label{eqn:MSE}
\end{align}
where we have used $b_N/a_N = \alpha = p(1+W)$ from \eqref{eqn:AandB}. The critical points $k^{\ast}$ of the mean squared error satisfy $\mu'(k^{\ast})=0$. After algebra, this condition reduces to
\begin{align}\label{eq:kstartranscendental}
    k^{\ast} =  \alpha + \log k^{\ast} - 3 + \frac{1}{\alpha + \log k^{\ast} -1}.
\end{align}
For $k^{\ast}$ fixed and $N\gg1$, we have $W\gg1$ and $\log k^{\ast} \ll W$. In Appendix \ref{app:kstar} we use these properties to deduce that the solution of \eqref{eq:kstartranscendental} has behavior
\begin{equation}\label{eq:kstarleading}
        k^\ast
    =\log N
    +(1-p)\log\log N
    +[p+\log (AC^p) -3]+o(1)\quad\text{as}\quad N\to\infty.
\end{equation}
Remarkably, we find that in the regime $N\gg1$, the optimal number of receptor events is $k^{\ast}\approx \log N$, independent of geometric factors which provide $\bigoh(1)$ corrections. In Appendix \ref{app:kstar} we explore the dependence of $k^{\ast}$ on these lower order corrections for some particular choices of boundary conditions.}

\begin{figure}
    \centering
    \subfigure[$\mu(k)$ against $k$ with optimal $k=k^{\ast}$.]{\includegraphics[width = 0.45\textwidth]{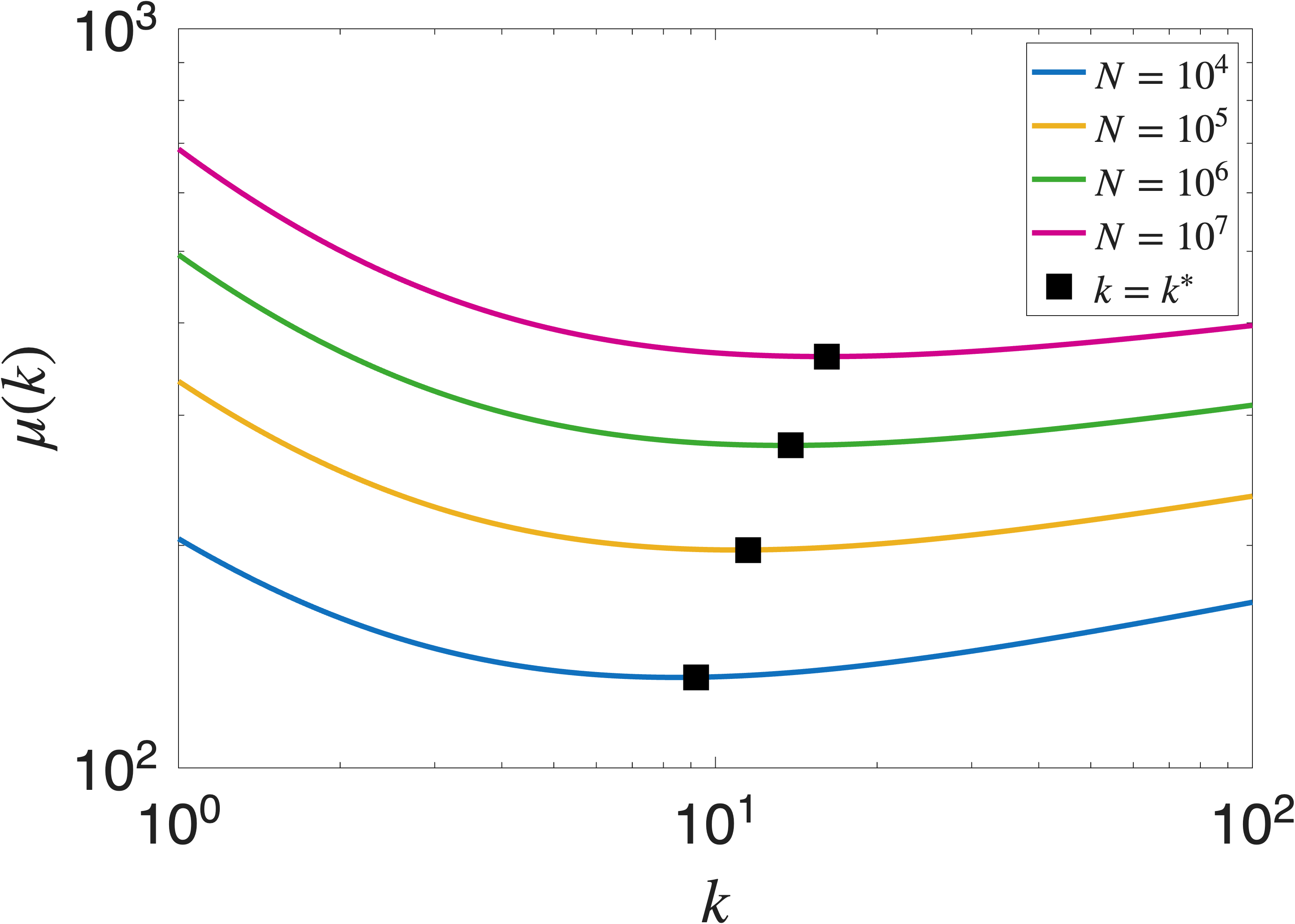}\label{fig:MSE_a}}\qquad
    \subfigure[The bias-variance trade-off against $k$.]{\includegraphics[width = 0.45\textwidth]{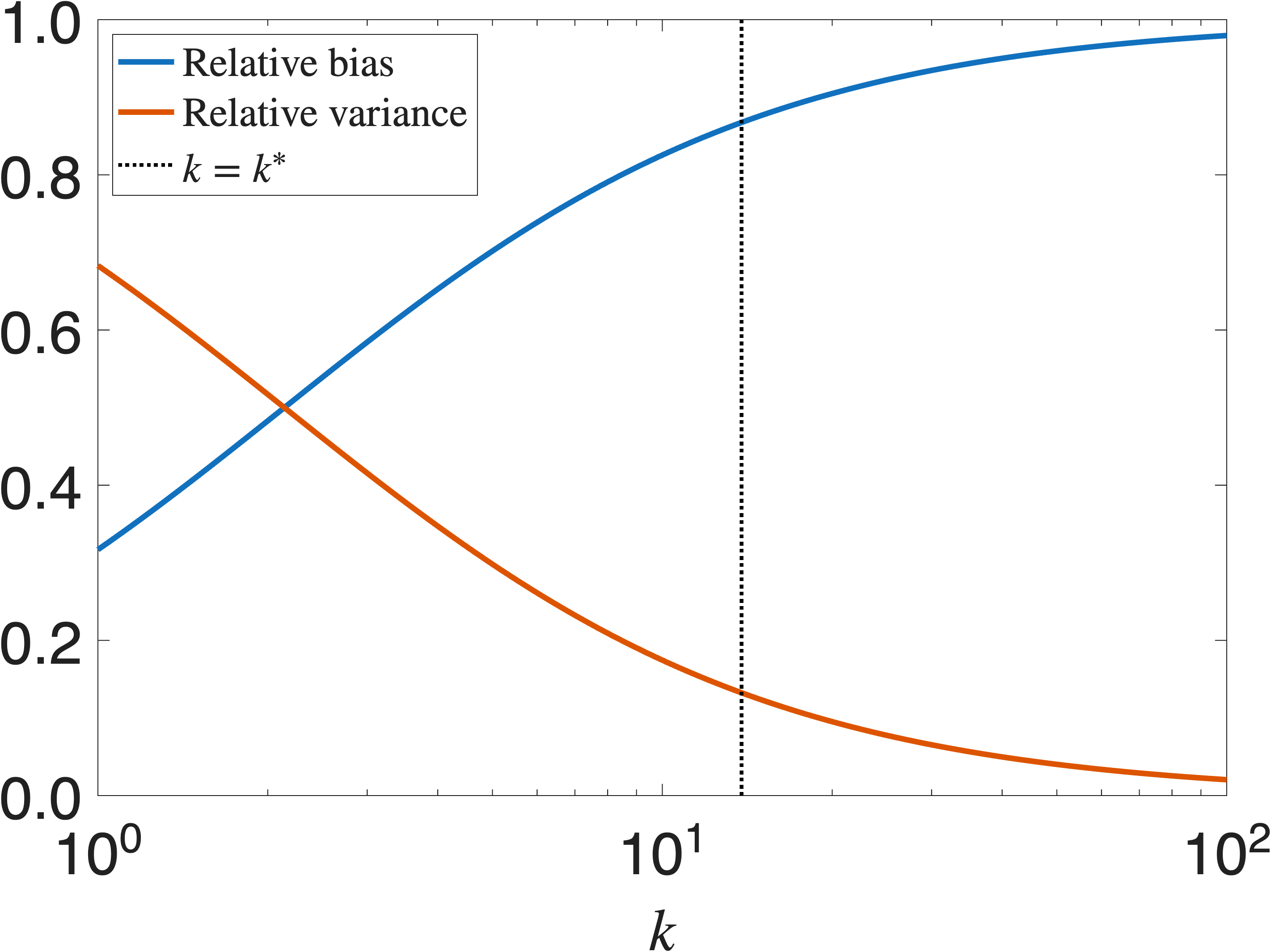}\label{fig:MSE_b}}
    \caption{\al{Bias-variance trade-off in the mean squared error (MSE) \eqref{eqn:MSE} as a function of impacts $k$ in the case $\alpha = \log N$. In panel (a) we plot $\mu(k)$ against $k$ for several values of $N$. We observe the MSE is minimized for $k^{\ast} \approx \log N$. In panel (b), we fix $N= 10^6$ and show the relative importance of bias and variance \eqref{eqn:relBiasVar} in the MSE as $k$ increases. For $k<k^{\ast}$, the MSE is decreasing as $k$ increases. Conversely,  for $k>k^{\ast}$, the MSE increases with the number of impacts.}\label{fig:MSE}}
\end{figure}

\al{In Fig.~\ref{fig:MSE} we display results for the case $\alpha\approx\log N$. In Fig.~\ref{fig:MSE_a} we show the minimization of the MSE \eqref{eqn:MSE} with $k$ for several values of $N\gg1$. The critical values $k^{\ast} \approx \log N$ accurately capture the threshold beyond which further impacts increase the MSE. In Fig.~\ref{fig:MSE_b}, we plot the relative bias and relative variance for fixed $N=10^6$. These quantities are defined as 
\begin{equation}\label{eqn:relBiasVar}
\text{Rel. bias}: \ \frac{\mathbb{E}\big[\rho_k^{\text{res}}\big]^2}{\mathbb{E}\big[\rho_k^{\text{res}}\big]^2+\text{Var}\big[\rho_k^{\text{res}}\big]}, \quad \text{Rel. variance:} \ \frac{\text{Var}\big[\rho_k^{\text{res}}\big]}{\mathbb{E}\big[\rho_k^{\text{res}}\big]^2 + \text{Var}\big[\rho_k^{\text{res}}\big]}.
\end{equation}
As seen in Fig.~\ref{fig:MSE_b}, the bias in the estimate increases with $k$ while the variance decreases with $k$. The vertical line $k= k^{\ast} \approx \log N $ is the critical value beyond which more impacts deteriorate the estimate. In Fig.~\ref{fig:MSE_a} the optimization landscape is seen to be relatively shallow indicating that for even $k\approx 10$, the MSE is very close to that of the optimal choice $k^{\ast}$ over several decades in $N$.
}

\section{Discussion}\label{sec:discussion}

In this work, we investigated the limits of single-cell directional sensing in the ultra-low chemoattractant regime, where receptor binding is discrete and cellular decisions occur long before steady state. A central mathematical contribution of this study is the derivation and validation of explicit asymptotic distributions for both the hitting times and hitting locations of diffusing signaling molecules at the cell surface. By coupling extreme first-passage time statistics with short-time angular localization, we obtained a joint statistical description of early receptor binding events that is valid in the biologically relevant pre-equilibrium regime.

The biological significance of this mathematical framework is that it provides a quantitative framework for comparing the directional information available to the cell under different hypothesized sensing mechanisms. Rather than assuming equilibrium gradients or time-averaged receptor occupancy, our results make it possible to directly analyze how much information is encoded in specific features of early binding data, such as first-arrival times, interarrival intervals, binding locations, or combinations thereof. This enables a systematic comparison of inference strategies, including statistically optimal estimators and simpler biologically plausible estimators, on equal footing. As an example, we found that the interarrival times conveyed much less information to the cell than the full arrival times, but exploiting the full arrival record requires nonlinear processing of the arrival times. 

Our analysis demonstrates that directional information is strongly front-loaded in early arrivals: molecules that bind soon after release are preferentially localized around the true source direction, while later arrivals lose directional bias. \al{Additionally, source distance is primarily encoded in the arrival times while the source direction is encoded in arrival locations.} As a result, accurate directional inference can be achieved from a small number of early binding events without the need for long-time averaging or high receptor occupancy. This provides a plausible mechanism that explains the experimentally observed rapid symmetry breaking and polarization responses that occur on timescales much shorter than those required to establish steady-state concentration profiles.

\al{A central simplification of our analysis is the treatment of the cell surface as a continuum of receptors, encoded in the homogenized Robin condition \eqref{eq:robin} with effective permeability $\kappa$. This description is justified when receptors are numerous and densely and uniformly distributed,
for which $\kappa\gg1$ and the surface is effectively absorbing
\cite{LindsayRSOS2023}. In this regime the fine-grained
arrangement of individual receptors is averaged out, and the arrival statistics depend only on the source geometry and the bulk diffusion.  Receptor polarization is characterized by a non-uniform surface density, possibly modulated by membrane curvature \cite{Endres2025}  or as an active element of the downstream chemotactic response \cite{NwogbagaCamley2024}. Capturing these effects requires moving beyond the homogenized boundary condition, for instance through discrete-trap or narrow-escape treatments \cite{LLM2020} or a matched-asymptotic analysis that resolves individual receptors while retaining the short-time structure in the survival probability. This is an important direction for extending the present results to more biologically realistic receptor configurations.}

Several other extensions of this framework remain open for further analysis. In particular, source detection in environments containing multiple chemoattractant sources presents new challenges, as early arrivals may reflect competing extreme events whose statistics overlap in time and space \cite{Camley24}. In congested environments, it is known that a community of cells can collaborate to better decipher environmental cues \cite{fancher2017,Camley2016,Camley2018,Copos2025}. Likewise, crowded or heterogeneous environments may modify first-passage statistics through obstruction, intermittency, or anomalous transport. Extending the joint hitting-time and hitting-location analysis to these settings would enable a quantitative assessment of how robust early-time directional sensing remains under realistic environmental complexity. Finally, an interesting avenue for future research will be to extend these results to more general cell shapes to further inform on the observed roles that aberrations to radial symmetry can play in cellular signaling processes \cite{CKL2025,LindsayBernoff2025,Camley2025,Copos2025,NwogbagaCamley2024}.

\section*{Acknowledgements}
AEL acknowledges support from the NSF under grant DMS-2052636. SDL acknowledges support from the NSF under grant DMS-2325258.

\section*{Data availability statement} \al{This manuscript has no associated experimental data. The code used to generate the figures and the numerical results are available from the corresponding author upon request.}

\appendix
\section{Extreme interarrival times are exponential}\label{app:Times_Diff}

As stated in equation \eqref{eqn_limitingT} (see also \cite[Thm.~4]{lawley2020dist}), the times $\{T_{1,N},\ldots,T_{k,N}\}$ obey the following convergence in distribution,
\begin{align}\label{eq:cd}
    \bigg(\frac{T_{1,N}-b_N}{a_N},\dots,\frac{T_{k,N}-b_N}{a_N}\bigg)
    \to_\dd(X_1,\dots,X_k)\quad\text{as }N\to\infty,
\end{align}
where the probability density function $f_{\textbf{X}^{(k)}}$ of $\textbf{X}^{(k)} = (X_1,\dots,X_k)$ is given in \eqref{eqn:dist_full}. The continuous mapping theorem ensures that continuous functions preserve convergence in distribution \cite{billingsley2013}. Hence, applying the following continuous function $g:\R^k\to\R^{k-1}$,
\begin{align*}
    g(x_1,\dots,x_k)
    =(x_2-x_1,\dots,x_k-x_{k-1}),
\end{align*}
to \eqref{eq:cd} yields the following convergence in distribution,
\begin{align*}
    \bigg(\frac{T_{2,N}-T_{1,N}}{a_N},\dots,\frac{T_{k,N}-T_{k-1,N}}{a_N}\bigg)
    \to_\dd(X_2-X_1,\dots,X_k-X_{k-1})\quad\text{as }N\to\infty.
\end{align*}

It thus remains to determine the joint distribution of $(X_2-X_1,\dots,X_k-X_{k-1})$. For $k=2$, we have that for $z\ge0$,
\begin{align*}
    \P(X_2-X_1>z)
    =\P(X_2>X_1+z)
    &=\int_{-\infty}^\infty\dd x_2\,e^{x_2}\exp(-e^{x_2})\int_{-\infty}^{x_2-z}\dd x_1\,e^{x_1}\\
    &=e^{-z}\int_{-\infty}^\infty\dd x_2\,e^{2x_2}\exp(-e^{x_2})=e^{-z}.
\end{align*}
Hence, $X_2-X_1$ is an exponential random variable with unit mean. 
For $k=3$, we have that for $z_1\ge0$, $z_2\ge0$,
\begin{align*}
    &\P(X_3-X_2>z_2,\, X_2-X_1>z_1)
    =\P(X_3>X_2+z_2,\, X_2>X_1+z_1)\\
    &\quad=\int_{-\infty}^\infty\dd x_3\,e^{x_3}\exp(-e^{x_3})\int_{-\infty}^{x_3-z_2}\dd x_2\,e^{x_2}\int_{-\infty}^{x_2-z_1}\dd x_1\,e^{x_1}\\
    &\quad=e^{-z_1}\int_{-\infty}^\infty\dd x_3\,e^{x_3}\exp(-e^{x_3})\int_{-\infty}^{x_3-z_2}\dd x_2\,e^{2x_2}\\
    &\quad=e^{-z_1-2z_2}\int_{-\infty}^\infty\dd x_3\,e^{3x_3}\exp(-e^{x_3})/2
    =e^{-z_1-2z_2}.
\end{align*}
Hence, $2(X_3-X_2)$ and $(X_2-X_1)$ are i.i.d. exponential random variables with unit means. For a general $k\ge2$, we have that for $z_1\ge0,\dots,z_{k-1}\ge0$,
\begin{align*}
    &\P(X_k-X_{k-1}>z_{k-1},\dots,X_2-X_1>z_1)
    =\P(X_k>X_{k-1}+z_{k-1},\dots,X_2>X_1+z_1)\\
    &=\int_{-\infty}^\infty \dd x_k\, e^{x_k}\exp(-e^{x_k})\int_{-\infty}^{x_k-z_{k-1}}\dd x_{k-1}\,e^{x_{k-1}}\int_{-\infty}^{x_{k-1}-z_{k-2}}\dd x_{k-2}\,e^{x_{k-2}}\cdots\int_{-\infty}^{x_2-z_{1}}\dd x_{1}\,e^{x_{1}}\\
    &=e^{-z_1-2z_2-3z_3-\cdots-(k-1)z_{k-1}}\int_{-\infty}^\infty \dd x_k\, e^{kx_k}\exp(-e^{x_k})/((k-1)!)\\
    &=e^{-z_1}e^{-2z_2}e^{-3z_3}\cdots e^{-(k-1)z_{k-1}}.
\end{align*}
Here, we have used that 
\begin{align*}
   &\int_{-\infty}^{x_k-z_{k-1}}\dd x_{k-1}\,e^{x_{k-1}}\int_{-\infty}^{x_{k-1}-z_{k-2}}\dd x_{k-2}\,e^{x_{k-2}}\cdots\int_{-\infty}^{x_2-z_{1}}\dd x_{1}\,e^{x_{1}}\\
   &\quad=e^{-z_1-2z_2-3z_3-\cdots-(k-1)z_{k-1}}e^{(k-1)x_k}/((k-1)!),
\end{align*}
which can be established by induction. We have also used that the change of variables $u=e^x$ yields
\begin{align*}
    \int_{-\infty}^\infty \dd x_k\, e^{kx_k}\exp(-e^{x_k})
    =\int_0^\infty \dd u\,u^{k-1}e^{-u}
    =\Gamma(k)=(k-1)!.
\end{align*}
We therefore conclude that
\begin{align*}
    \big\{i(X_{i+1}-X_i)\big\}_{i=1}^{k-1}
    =\Big\{X_2-X_1,2(X_3-X_2),3(X_4-X_3),\dots,(k-1)(X_k-X_{k-1})\Big\}
\end{align*}
are i.i.d. exponential random variables with unit means.

\subsection{$X_1$ and $X_2-X_1$ are dependent}

Suppose we want to sample $(X_1,\dots,X_k)$ from the pdf in \eqref{eqn:dist_full}. As a first attempt, one might first sample $X_1\sim\text{Gumbel}(0,1)$ and then let $X_2=X_1+E$ where $E$ is an exponential random variable independent of $X_1$. However, this sampling method assumes that $X_1$ and $X_2-X_1$ are independent. However, this is false. In particular, while all the \textit{inter}arrival times are independent, the first arrival time and the first interarrival time are \textit{not} independent. This is intuitive; if the first particle takes a long time to hit the target (i.e.\ $X_1$ is large), then it seems likely that the first interarrival time will be small (i.e.\ $X_2-X_1$ is small). To see that $X_1$ and $X_2-X_1$ are dependent, we compute 
\begin{align*}
    \mathbb{E}[X_1(X_2-X_1)]
    &=\int\int x_1(x_2-x_1)f(x_1,x_2)\,\dd x_1\,\dd x_2
    =-1-\gamma_e\\
    &\neq \mathbb{E}[X_1]\E[X_2-X_1]=(-\gamma_e)\times1=-\gamma_e.
\end{align*}

\section{\sdl{Next order asymptotics of optimal $k^\ast$}}\label{app:kstar}

\sdl{We now derive the next order asymptotics of the optimal $k^\ast$ derived in section~\ref{sec:Source_Extreme}. Recall from \eqref{eq:kstarleading} that $k^\ast\sim\alpha=p(1+W)\sim\log N$ as $N\to\infty$ where $W=W_0(z)$ and $z=(C/p)(AN)^{1/p}$ for $p>0$. Setting $L=\log z$, the LambertW function has the following large argument expansion \cite{corless1996},
\begin{align}\label{eq:Wlarge}
    W_0(z)
    =L-\log L+O\Big(\frac{\log L}{L}\Big)\quad\text{as }L\to\infty.
\end{align}
Since
\begin{align*}
    L=\log z=\log((C/p)(AN)^{1/p})
    =\frac{1}{p}\log N+\frac{1}{p}\log A+\log(C/p)\sim\frac{1}{p}\log N\quad\text{as }N\to\infty,
\end{align*}
we have that
\begin{align*}
    \log L
    =\log\log N-\log p
    +O\Big(\frac{1}{\log N}\Big)\quad\text{as }N\to\infty.
\end{align*}
Combining this with \eqref{eq:Wlarge} yields
\begin{align}\label{eq:kstarnext}
    \alpha
    =p(1+W)
    =\log N-p\log\log N+p+\log A+p\log C+o(1)\quad\text{as }N\to\infty,
\end{align}
which, after combining with the relation $k^\ast=\alpha+\log k^\ast-3+1/(\alpha+\log k^\ast-1)$ in \eqref{eq:kstartranscendental}, yields
\begin{align*}
    k^\ast
    =\log N
    +(1-p)\log\log N
    +[p+\log A+p\log C-3]+o(1)\quad\text{as }N\to\infty.
\end{align*}
It is straightforward to check that this estimate is self-consistent with \eqref{eq:kstartranscendental}. Hence, the geometry-dependent constants $A$ and $C$ affect the optimal $k^\ast$ through an additive constant. For the case of a circular cell with Dirichlet boundary conditions, the constant in \eqref{eq:constants} ($\kappa=\infty$) yields
\begin{align*}
    k^\ast_{\text{Dir}}
    =\log N
    +\tfrac{1}{2}\log\log N-\tfrac{1}{2}\log(\pi R)-5/2+o(1)\quad\text{as }N\to\infty.
\end{align*}
Notice that all of the factors of the diffusivity $D$ have canceled. Hence, in accord with the rest of our results, the diffusivity controls the timescale of when particles arrive, but it does not affect the optimal number to use for source inference. Notice also that farther sources (larger $R$) decreases the number of arrivals worth retaining, which has the interpretation that directional information degrades faster with respect to the ordered arrivals for more distant sources.
}

\bibliography{library.bib}

@article{NwogbagaCamley2024,
  title = {Cell shape and orientation control galvanotactic accuracy},
  author = {Nwogbaga, Ifunanya and Camley, Brian A.},
  journal = {Soft Matter},
  year = {2024},
  volume = {20},
  pages = {8866--8887},
  doi = {10.1039/D4SM00952E}
}

@article{Endres2025,
  title = {Local Clustering and Global Spreading of Receptors for Optimal Spatial Gradient Sensing},
  author = {Alonso, Albert and Endres, Robert G. and Kirkegaard, Julius B.},
  journal = {Phys. Rev. Lett.},
  volume = {134},
  issue = {15},
  pages = {158401},
  numpages = {6},
  year = {2025},
  month = {Apr},
  publisher = {American Physical Society},
  doi = {10.1103/PhysRevLett.134.158401},
  url = {https://link.aps.org/doi/10.1103/PhysRevLett.134.158401}
}

@article{Mogilner2023,
author = {Sun, Yaohui and Reid, Brian and Zhang, Yan and Zhu, Kan and Ferreira, Fernando and Estrada, Alejandro and Sun, Yuxin and Draper, Bruce W. and Yue, Haicen and Copos, Calina and Lin, Francis and Bernadskaya, YelenaY. and Zhao, Min and Mogilner, Alex},
title = {Electric field–guided collective motility initiation of large epidermal cell groups},
journal = {Molecular Biology of the Cell},
volume = {34},
number = {5},
pages = {ar48},
year = {2023},
doi = {10.1091/mbc.E22-09-0391},
    note ={PMID: 36989037},
URL = {https://doi.org/10.1091/mbc.E22-09-0391}
}

@article{Camley24,
	author = {Kashyap, Aparajita and Wang, Wei and Camley, Brian A},
	journal = {Biophysical Journal},
	month = {2026/02/17},
	number = {10},
	pages = {1184--1194},
	title = {Trade-offs in concentration sensing in dynamic environments},
	volume = {123},
	year = {2024}}

@Book{abramowitzstegun,
 author    = "Milton {Abramowitz} and Irene A. {Stegun}",
 title     = "Handbook of Mathematical Functions with
              Formulas, Graphs, and Mathematical Tables",
 publisher = "Dover",
 year      =  1964,
 address   = "New York City",
 edition   = "ninth Dover printing, tenth GPO printing"
}

@article{CHERRY2025,
title = {Boundary integral methods for particle diffusion in complex geometries: Shielding, confinement, and escape},
journal = {Journal of Computational Physics},
volume = {534},
pages = {114032},
year = {2025},
issn = {0021-9991},
doi = {https://doi.org/10.1016/j.jcp.2025.114032},
url = {https://www.sciencedirect.com/science/article/pii/S0021999125003158},
author = {Jesse Cherry and Alan E. Lindsay and Bryan D. Quaife}
}

@article{genchi2,
author = {Abhranil Das},
title = {New methods to compute the generalized chi-square distribution},
journal = {Journal of Statistical Computation and Simulation},
volume = {95},
number = {12},
pages = {2608--2642},
year = {2025},
publisher = {Taylor \& Francis},
doi = {10.1080/00949655.2025.2501401},
}

@article{GPCR_size,
author ="Odoemelam, Chiemela S. and Percival, Benita and Wallis, Helen and Chang, Ming-Wei and Ahmad, Zeeshan and Scholey, Dawn and Burton, Emily and Williams, Ian H. and Kamerlin, Caroline Lynn and Wilson, Philippe B.",
title  ="G-Protein coupled receptors: structure and function in drug discovery",
journal  ="RSC Adv.",
year  ="2020",
volume  ="10",
issue  ="60",
pages  ="36337-36348",
publisher  ="The Royal Society of Chemistry",
doi  ="10.1039/D0RA08003A",
url  ="http://dx.doi.org/10.1039/D0RA08003A"}

@article{Ismael2016,
author = {Ismael, Amber and Tian, Wei and Waszczak, Nicholas and Wang, Xin and Cao, Youfang and Suchkov, Dmitry and Bar, Eli and Metodiev, Metodi V. and Liang, Jie and Arkowitz, Robert A. and Stone, David E.},
issn = {19379145},
journal = {Sci. Signal.},
number = {423},
pages = {1--17},
title = {{G$\beta$ promotes pheromone receptor polarization and yeast chemotropism by inhibiting receptor phosphorylation}},
volume = {9},
year = {2016}
}

@article{Lakhani2017,
author = {Lakhani, Vinal and Elston, Timothy C.},
isbn = {1111111111},
issn = {15537358},
journal = {PLoS Comput. Biol.},
number = {2},
pages = {1--30},
title = {{Testing the limits of gradient sensing}},
volume = {13},
year = {2017}
}

@article{LawleyJohnson2023,
	author = {Lawley, Sean D. and Johnson, Joshua},
	journal = {Journal of Mathematical Biology},
	number = {6},
	pages = {90},
	title = {Slowest first passage times, redundancy, and menopause timing},
	volume = {86},
	year = {2023}}

@article{Linn2024,
  title = {Fast decisions reflect biases; slow decisions do not},
  author = {Linn, S and Lawley, S. D. and Karamched, B. R. and Kilpatrick, Z. P. and Josi\'{c}, K.},
  journal = {Phys. Rev. E},
  volume = {110},
  issue = {2},
  pages = {024305},
  numpages = {18},
  year = {2024},
  month = {Aug},
  publisher = {American Physical Society},
  doi = {10.1103/PhysRevE.110.024305},
  url = {https://link.aps.org/doi/10.1103/PhysRevE.110.024305}
}

@article{Lew2019,
	author = {Henderson, Nicholas T. AND Pablo, Michael AND Ghose, Debraj AND Clark-Cotton, Manuella R. AND Zyla, Trevin R. AND Nolen, James AND Elston, Timothy C. AND Lew, Daniel J.},
	journal = {PLOS Biology},
	month = {10},
	number = {10},
	pages = {1-35},
	title = {Ratiometric GPCR signaling enables directional sensing in yeast},
	volume = {17},
	year = {2019}}

@article{Parent1999,
	author = {Parent, C A and Devreotes, P N},
	journal = {Science},
	month = {Apr},
	number = {5415},
	pages = {765--770},
	title = {A cell's sense of direction.},
	volume = {284},
	year = {1999}}

@article{Servant2000,
	author = {Servant, G and Weiner, O D and Herzmark, P and Balla, T and Sedat, J W and Bourne, H R},
	journal = {Science},
	month = {Feb},
	number = {5455},
	pages = {1037--1040},
	title = {Polarization of chemoattractant receptor signaling during neutrophil chemotaxis.},
	volume = {287},
	year = {2000}}

@article{Orion1999,
	author = {Weiner, Orion D. and Servant, Guy and Welch, Matthew D. and Mitchison, Timothy J. and Sedat, John W. and Bourne, Henry R.},
	journal = {Nature Cell Biology},
	number = {2},
	pages = {75--81},
	title = {Spatial control of actin polymerization during neutrophil chemotaxis},
	volume = {1},
	year = {1999}}

@article{civciristov2018,
  title={Preassembled GPCR signaling complexes mediate distinct cellular responses to ultralow ligand concentrations},
  author={Civciristov, Srgjan and Ellisdon, Andrew M and Suderman, Ryan and Pon, Cindy K and Evans, Bronwyn A and Kleifeld, Oded and Charlton, Steven J and Hlavacek, William S and Canals, Meritxell and Halls, Michelle L},
  journal={Science signaling},
  volume={11},
  number={551},
  pages={eaan1188},
  year={2018},
  publisher={American Association for the Advancement of Science}
}

@article{Orion2002,
	author = {Weiner, Orion D},
	journal = {Curr Opin Cell Biol},
	month = {Apr},
	number = {2},
	pages = {196--202},
	title = {Regulation of cell polarity during eukaryotic chemotaxis: the chemotactic compass.},
	volume = {14},
	year = {2002}}

@Inbook{Millius2009,
author="Millius, Arthur and Weiner, Orion D.",
editor="Jin, Tian and Hereld, Dale",
title="Chemotaxis in Neutrophil-Like HL-60 Cells",
bookTitle="Chemotaxis: Methods and Protocols",
year="2009",
publisher="Humana Press",
address="Totowa, NJ",
pages="167--177",
isbn="978-1-60761-198-1"
}

@article{Levchenko2002,
author = {Levchenko, Andre and Iglesias, Pablo A.},
doi = {10.1016/S0006-3495(02)75373-3},
isbn = {4105166026},
issn = {00063495},
journal = {Biophys. J.},
number = {1},
pages = {50--63},
publisher = {Elsevier},
title = {{Models of eukaryotic gradient sensing: Application to chemotaxis of amoebae and neutrophils}},
url = {http://dx.doi.org/10.1016/S0006-3495(02)75373-3},
volume = {82},
year = {2002}
}

@article{CALABRESE2021,
title = {Ultra low doses and biological amplification: Approaching Avogadro’s number},
journal = {Pharmacological Research},
volume = {170},
pages = {105738},
year = {2021},
issn = {1043-6618},
doi = {https://doi.org/10.1016/j.phrs.2021.105738},
url = {https://www.sciencedirect.com/science/article/pii/S1043661821003224},
author = {Edward J. Calabrese and James Giordano}
}

@article{Boltz2022,
	author = {Boltz, Horst-Holger and Sirbu, Alexei and Stelzer, Nina and de Lanerolle, Primal and Winkelmann, Stefanie and Annibale, Paolo},
	journal = {Cells},
	month = {May},
	number = {10},
	title = {The Impact of Membrane Protein Diffusion on GPCR Signaling.},
	volume = {11},
	year = {2022}}

@article{CKL2025,
 title={The effect of target orientation on the mean first passage time of a Brownian particle to a small elliptical absorber},
 DOI={10.1017/S095679252510020X},
 journal={European Journal of Applied Mathematics},
 author={Chakraborty, Sanchita and Kolokolnikov, Theodore and Lindsay, Alan E.},
 year={2025},
 pages={1–25}
 }

@article{Civciristov2019,
	author = {Civciristov, Srgjan and Halls, Michelle L},
	journal = {Br J Pharmacol},
	month = {Jul},
	number = {14},
	pages = {2382--2401},
	title = {Signalling in response to sub-picomolar concentrations of active compounds: Pushing the boundaries of GPCR sensitivity.},
	volume = {176},
	year = {2019}}

@article{Camley2016,
	author = {Camley, Brian A. and Zimmermann, Juliane and Levine, Herbert and Rappel, Wouter Jan},
	journal = {PLoS Comput. Biol.},
	number = {7},
	pages = {1--28},
	title = {{Collective Signal Processing in Cluster Chemotaxis: Roles of Adaptation, Amplification, and Co-attraction in Collective Guidance}},
	volume = {12},
	year = {2016}}

@article{Camley2018,
	author = {Camley, Brian A},
	journal = {J. Phys. Condens. Matter},
	month = {jun},
	number = {22},
	pages = {223001},
	title = {{Collective gradient sensing and chemotaxis: modeling and recent developments}},
	volume = {30},
	year = {2018}}

@Article{moth2020,
AUTHOR = {Stepien, Tracy L. and Zmurchok, Cole and Hengenius, James B. and Caja Rivera, Rocío Marilyn and D’Orsogna, Maria R. and Lindsay, Alan E.},
TITLE = {Moth Mating: Modeling Female Pheromone Calling and Male Navigational Strategies to Optimize Reproductive Success},
JOURNAL = {Applied Sciences},
VOLUME = {10},
YEAR = {2020},
NUMBER = {18},
ARTICLE-NUMBER = {6543},
URL = {https://www.mdpi.com/2076-3417/10/18/6543},
ISSN = {2076-3417},
DOI = {10.3390/app10186543}
}

@article{lawley2020networks,
  title={Extreme first-passage times for random walks on networks},
  author={Lawley, Sean D},
  journal={Physical Review E},
  volume={102},
  number={6},
  pages={062118},
  year={2020},
  publisher={APS}
}

@article{lawley2020dist,
  title={Distribution of extreme first passage times of diffusion},
  author={Lawley, S D},
  journal={Journal of Mathematical Biology},
  year={2020},
  doi={10.1007/s00285-020-01496-9},
  url = {http://dx.doi.org/10.1007/s00285-020-01496-9}
}

@book{billingsley2013,
  title={Convergence of Probability Measures},
  author={Billingsley, Patrick},
  isbn={9781118625965},
  series={Wiley Series in Probability and Statistics},
  year={2013},
  publisher={John Wiley \& Sons},
  edition={2nd}
}

@article{corless1996,
  title={On the {LambertW} function},
  author={Corless, RM and Gonnet, GH and Hare, DEG and Jeffrey, DJ and Knuth, DE},
  journal={Advances in Computational mathematics},
  volume={5},
  number={1},
  pages={329--359},
  year={1996},
  publisher={Springer}
}

@article{LindsayRSOS2023,
	author = {Lindsay, Alan E. and Bernoff, Andrew J. and Navarro Hern{\'a}ndez, Adri{\'a}n},
	journal = {Royal Society Open Science},
	month = {04},
	number = {4},
	pages = {221619},
	title = {Short-time diffusive fluxes over membrane receptors yields the direction of a signalling source},
	volume = {10},
	year = {2023}}

@article{newby2016,
  title={First-passage time to clear the way for receptor-ligand binding in a crowded environment},
  author={Newby, J and Allard, J},
  journal={Phys Rev Lett},
  volume={116},
  number={12},
  pages={128101},
  year={2016},
  publisher={APS}
}

@article{Dobramysl2018b,
	author = {U. Dobramysl and D. Holcman},
	journal = {Journal of Computational Physics},
	pages = {22-36},
	title = {Mixed analytical-stochastic simulation method for the recovery of a Brownian gradient source from probability fluxes to small windows},
	volume = {355},
	year = {2018}}

@article{Dobramysl2018a,
	author = {Dobramysl, Ulrich and Holcman, David},
	journal = {Scientific Reports},
	number = {1},
	pages = {941},
	title = {Reconstructing the gradient source position from steady-state fluxes to small receptors},
	volume = {8},
	year = {2018}}

@article{LLM2020,
  title = {Receptor Organization Determines the Limits of Single-Cell Source Location Detection},
  author = {Lawley, Sean D. and Lindsay, Alan E. and Miles, Christopher E.},
  journal = {Phys. Rev. Lett.},
  volume = {125},
  issue = {1},
  pages = {018102},
  numpages = {6},
  year = {2020},
  month = {Jul},
  publisher = {American Physical Society},
  doi = {10.1103/PhysRevLett.125.018102},
  url = {https://link.aps.org/doi/10.1103/PhysRevLett.125.018102}
}

@article{lindsay2015,
	title = {Narrow escape problem with a mixed trap and the effect of orientation},
	volume = {91},
	issn = {1539-3755, 1550-2376},
	url = {http://link.aps.org/doi/10.1103/PhysRevE.91.032111},
	doi = {10.1103/PhysRevE.91.032111},
	language = {en},
	number = {3},
	urldate = {2015-08-10},
	journal = {Phys Rev E},
	author = {Lindsay, A. E. and Kolokolnikov, T. and Tzou, J. C.},
	month = mar,
	year = {2015}
}

@article{fancher2017,
  title={Fundamental limits to collective concentration sensing in cell populations},
  author={Fancher, S and Mugler, A},
  journal={Phys Rev Lett},
  volume={118},
  number={7},
  pages={078101},
  year={2017},
  publisher={APS}
}

@article{Brennan:2026,
	author = {Brennan, Molly and Yeo, Edwina F. and Pearce, Philip and Dalwadi, Mohit P.},
	journal = {Proceedings of the Royal Society A: Mathematical, Physical and Engineering Sciences},
	month = {2/17/2026},
	number = {2331},
	pages = {20250703},
	title = {Effective permeability conditions for diffusive transport through impermeable membranes with gaps},
	volume = {482},
	year = {2026}}

@article{Mugler2016,
  title={Limits to the precision of gradient sensing with spatial communication and temporal integration},
  author={Mugler, A and Levchenko, A and Nemenman, I},
  journal={Proc Natl Acad Sci},
  volume={113},
  number={6},
  pages={E689--E695},
  year={2016},
  publisher={National Acad Sciences}
}

@article{aquino2016,
  title={Know the single-receptor sensing limit? Think again},
  author={Aquino, G and Wingreen, N S and Endres, R G},
  journal={J Stat Phys},
  volume={162},
  number={5},
  pages={1353--1364},
  year={2016},
  publisher={Springer}
}

@article{MacLaurin2025,
	author = {MacLaurin, James and Newby, Jay},
	journal = {SIAM Journal on Applied Mathematics},
	number = {1},
	pages = {109-142},
	title = {Extreme First Passage Times for Populations of Identical Rare Events},
	volume = {85},
	year = {2025}}

@article{Linn_2022,
	author = {Linn, Samantha and Lawley, Sean D},
	journal = {Journal of Physics A: Mathematical and Theoretical},
	month = {aug},
	number = {34},
	pages = {345002},
	title = {Extreme hitting probabilities for diffusion},
	volume = {55},
	year = {2022}}

@article{EndresWingreen2009,
  title = {Maximum Likelihood and the Single Receptor},
  author = {Endres, Robert G. and Wingreen, Ned S.},
  journal = {Phys. Rev. Lett.},
  volume = {103},
  issue = {15},
  pages = {158101},
  numpages = {4},
  year = {2009},
  month = {Oct},
  publisher = {American Physical Society},
  doi = {10.1103/PhysRevLett.103.158101},
  url = {https://link.aps.org/doi/10.1103/PhysRevLett.103.158101}
}

@article{Copos2025,
	author = {Calina Copos and Yao-Hui Sun and Kan Zhu and Yan Zhang and Brian Reid and Bruce Draper and Francis Lin and Haicen Yue and Yelena Bernadskaya and Min Zhao and Alex Mogilner},
	journal = {Proceedings of the National Academy of Sciences},
	number = {21},
	pages = {e2416440122},
	title = {Galvanotactic directionality of cell groups depends on group size},
	volume = {122},
	year = {2025}}

@article{Tung2025,
	author = {Tung, Hwai-Ray and Lawley, Sean D.},
	journal = {SIAM Journal on Applied Mathematics},
	number = {5},
	pages = {2145-2166},
	title = {First Passage Times with Fast Immigration},
	volume = {85},
	year = {2025}}

@book{vanderVaart1998,
  author    = {A. W. van der Vaart},
  title     = {Asymptotic Statistics},
  series    = {Cambridge Series in Statistical and Probabilistic Mathematics},
  volume    = {3},
  publisher = {Cambridge University Press},
  address   = {Cambridge},
  year      = {1998},
  isbn      = {978-0521496032}
}

@article{LM2023,
	author = {Morgan, Jonathan AND Lindsay, Alan E.},
	journal = {PLOS Computational Biology},
	month = {08},
	number = {8},
	pages = {1-17},
	title = {Modulation of antigen discrimination by duration of immune contacts in a kinetic proofreading model of T cell activation with extreme statistics},
	volume = {19},
	year = {2023}}

@article{Camley2025,
	author = {Kaiyrbekov, Kurmanbek and Camley, Brian A},
	journal = {PLoS One},
	number = {6},
	pages = {e0325800},
	title = {Does nematic order allow groups of elongated cells to sense electric fields better?},
	volume = {20},
	year = {2025}}

@article{LindBJ2023,
	author = {Bernoff, Andrew J. and Jilkine, Alexandra and Navarro Hern{\'a}ndez, Adri{\'a}n and Lindsay, Alan E.},
	journal = {Biophysical Journal},
	month = {2026/03/10},
	number = {15},
	pages = {3108--3116},
	title = {Single-cell directional sensing from just a few receptor binding events},
	volume = {122},
	year = {2023}}

@article{LindsayBernoff2025,
	author = {Bernoff, Andrew J. and Lindsay, Alan E.},
	journal = {Royal Society Open Science},
	month = {02},
	number = {2},
	pages = {241033},
	title = {Kinetic Monte Carlo methods for three-dimensional diffusive capture problems in exterior domains},
	volume = {12},
	year = {2025}}

@article{berg1977,
  title={Physics of chemoreception},
  author={Berg, Howard C and Purcell, Edward M},
  journal={Biophys J},
  volume={20},
  number={2},
  pages={193--219},
  year={1977},
  publisher={Elsevier}
}

\end{document}